\documentclass[english]{aa}
\usepackage{epsf,amsfonts,amssymb,graphicx,fancyheadings,caption}
\usepackage{comment}
\usepackage{rotating}
\usepackage{natbib}
\usepackage{babel}
\usepackage{txfonts}
\usepackage{float} 
\bibpunct{(}{)}{;}{a}{}{,}

\begin{document}

\title{The ongoing pursuit of R Coronae Borealis stars: ASAS-3 survey strikes again.}

\author{
P.~Tisserand\inst{1,2},
G.C.~Clayton\inst{3},
D.L.~Welch\inst{4},
B.~Pilecki\inst{5,6},
L.~Wyrzykowski\inst{6,7}, and
D.~Kilkenny\inst{8}
}

\institute{
Research School of Astronomy and Astrophysics, Australian National University, Cotter Rd, Weston Creek, ACT 2611, Australia \and
ARC Centre of Excellence for All-sky Astrophysics (CAASTRO) \and
Department of Physics \& Astronomy, Louisiana State University, Baton Rouge, LA 70803, USA \and
Department of Physics \& Astronomy, McMaster University, Hamilton, Ontario, L8S 4M1, Canada \and
Universidad de Concepci\'{o}n, Departamento de Astronom\'{i}a, Casilla 160-C, Concepci\'{o}n, Chile \and
Warsaw University Astronomical Observatory, Al. Ujazdowskie 4, 00-478 Warszawa, Poland \and
Institute of Astronomy, University of Cambridge, Madingley Road, Cambridge CB3 0HA, England \and
Department of Physics, University of the Western Cape, Private Bag X17, Bellville 7535, South Africa
}

\offprints{Patrick Tisserand; \email{tisserand@mso.anu.edu.au}}

\date{Received ; Accepted}

\abstract {R Coronae Borealis stars (RCBs) are rare, hydrogen-deficient, carbon-rich supergiant variable stars that are likely the evolved merger products of pairs of CO and He white dwarfs. Only 55 RCB stars have been found in our galaxy and their distribution on the sky is weighted heavily by microlensing survey field positions. A less-biased wide-area survey would provide the ability to test competing evolutionary scenarios, understand the population or populations that produce RCBs and constraint their formation rate.}
{The ASAS-3 survey monitored the sky south of declination +28 deg between 2000 and 2010 to a limiting magnitude of V = 14. We searched ASAS-3 for RCB variables using a number of different methods to ensure that the probability of RCB detection was as high as possible and to reduce selection biases based on luminosity, temperature, dust production activity and shell brightness.}
{Candidates whose light curves were visually inspected were pre-selected based on their infrared (IR) excesses due to warm dust in their circumstellar shells using the WISE and/or 2MASS catalogues. Criteria on light curve variability were also applied when necessary to minimize the number of objects. Initially, we searched for RCB stars among the ASAS-3 ACVS1.1 variable star catalogue, then among the entire ASAS-3 south source catalogue, and finally directly interrogated the light curve database for objects that were not catalogued in either of those. We then acquired spectra of 104 stars to determine their real nature using the SSO/WiFeS spectrograph.} 
{We report 21 newly-discovered RCB stars and 2 new DY Per stars. Two previously suspected RCB candidates were also spectroscopically confirmed. Our methods allowed us to extend our detection efficiency to fainter magnitudes that would not have been easily accessible to discovery techniques based on light curve variability. The overall detection efficiency is about 90\% for RCBs with maximum light brighter than $V\sim13$.} 
{With these new discoveries, 76 RCBs are now known in our Galaxy and 22 in the Magellanic Clouds. This growing sample is of great value to constrain the peculiar and disparate atmosphere composition of RCBs. Most importantly, we show that the spatial distribution and apparent magnitudes of Galactic RCB stars is consistent with RCBs being part of the Galactic bulge population.}


\authorrunning{P. Tisserand et al.}
\titlerunning{New R Coronae Borealis stars discovered in the ASAS-3 south dataset.}

\maketitle

\section{Introduction \label{sec_intro}}

The merger of two white dwarfs (WDs) is a rare astrophysical phenomenon that concludes the life of a close binary system. It is the source of a very interesting new class of stars or phenomena that are much studied nowadays. The merger of two Helium-WDs, a low mass system ($\mathrm{M_{Tot}}\lesssim 0.6 \mathrm{M_{\odot}}$), results in hot sub-dwarf type (sdB and sdO) stars, that explain the excess of UV emission observed in elliptical galaxies \citep{2008ASPC..392...15P}. At the other end of the mass scale, the merger of two CO-WDs ($\mathrm{M_{Tot}}\gtrsim 1.4 \mathrm{M_{\odot}}$) can produce one of the most powerful cataclysmic events known in our Universe, a Type Ia Supernova  \citep{2010ApJ...714L..52S,2012ApJ...747L..10P} or result in a neutron star after a gravitational collapse. In the middle range, when the total mass of the system does not exceed the Chandrasekhar mass limit, i.e., the merger of an He-WD with a CO-WD, a new class of stars may be created. R Coronae Borealis stars (RCBs) are the favoured candidates \citep{1984ApJ...277..355W,2011MNRAS.414.3599J,2012ApJ...748...35S,2012ApJ...757...76S}.

RCBs are hydrogen-deficient and carbon-rich supergiant stars. They are rare, with only 53 known in our galaxy \citep{1996PASP..108..225C,2008A&A...481..673T,2012JAVSO..40..539C}, and their atmosphere is made of 99\% helium. The  	
scarcity and unusual atmospheric composition of RCBs suggest that they correspond either to a brief phase of stellar evolution or an uncommon evolutionary path. Two major evolutionary scenarios have been suggested to explain their origin: the Double Degenerate (DD) scenario, corresponding to the merger of two WDs, and the final helium Shell Flash (FF) scenario \citep{1996ApJ...456..750I,1990ASPC...11..549R} which involves the expansion of a star, on the verge of becoming a white dwarf, to supergiant size (a late thermal pulse). The DD model has been strongly supported by the observations of an $^{18}$O overabundance in hydrogen-deficient carbon (HdC) and cool RCB stars \citep{2007ApJ...662.1220C,2010ApJ...714..144G}, 500 times higher than the $^{18}$O solar abundance \citep{1989GeCoA..53..197A}, but also the surface abundance anomalies for a few elements, fluorine in particular \citep{2008ApJ...674.1068P,2011MNRAS.414.3599J} and the evolutionary time scale \citep{2002MNRAS.333..121S}. On the other hand, nebulae were detected around four RCB stars \citep{2011ApJ...743...44C}, which indicates that a fraction of RCBs could still result from the FF scenario.

RCB stars are of spectral type K to F with an effective temperature ranging mostly between 4000 and 8000 K \citep{2000A&A...353..287A,2012A&A...539A..51T}, with a few exceptions being hotter than 10000 K (V348 Sgr, DY Cen,  MV Sgr and HV 2671) \citep{2002AJ....123.3387D}. Interestingly, \citet{2012ApJ...760L...3K} have shown recently that a member of this latter group, DY Cen, is in a binary system. This is the first star classified as an RCB star that is found to have a companion. Their study and the peculiar abundance of DY Cen indicates that is has certainly not followed the same evolution path than classical RCB, but has possibly gone though a true common-envelope system in the past 100 years. 

It is necessary to increase the known sample of RCB stars to answer our questions about their origin and population. With an absolute magnitude ranging from $-5\leqslant \mathrm{M_V}\leqslant -3.5$ \citep{2001ApJ...554..298A,2009A&A...501..985T} and a bright circumstellar shell, it is possible nowadays to detect the vast majority of RCB stars in the Milky Way and the Magellanic Clouds by using the datasets accumulated by ongoing surveys. In the favoured DD scenario, such a volume-limited search would allow us to determine the rate of He-WD + CO-WD mergers and therefore give constraints on population synthesis models \citep{2009ApJ...699.2026R}. 
Indeed, current models estimate a He-WD + CO-WD merger birthrate between $\sim\mathrm{10^{-3}}$ and $\sim\mathrm{5\times10^{-3}}$ per year \citep{2001A&A...365..491N,2009ApJ...699.2026R}, which would correspond to about 100 to 500 RCB stars existing nowadays in our Galaxy, using an RCB phase lifetime of about $10^5$ years, as predicted by theoretical evolution models made by \citet{2002MNRAS.333..121S}. Another interesting outcome of our endeavour to increase the number of known RCB stars is that we can constrain simulations of merging events. Indeed, the detailed study of RCBs' peculiar and diverse atmospheric composition would give further information to He-WD + CO-WD merger models with nucleosynthesis \citep{2011MNRAS.414.3599J,2012ApJ...757...76S}, which could in turn improve models of higher mass WD mergers that can cause transient type Ia Supernovae events.

The spectroscopic follow-up of candidates is the final step required to confirm them as RCB stars, as one needs to confirm the hydrogen deficiency and high C/O ratio of the atmosphere and also, in most cases, the low isotopic $^{13}$C/$^{12}$C ratio and high $^{18}$O/$^{16}$O ratio. Two different techniques are used to select candidates. One can select objects that possess a hot circumstellar shell using IR datasets \citep{2012A&A...539A..51T}, or apply algorithms to detect aperiodic fast and large declines in brightness among millions of light curves obtained in the visible region of the spectrum \citep{2001ApJ...554..298A,2004A&A...424..245T,2008A&A...481..673T,2012ApJ...755...98M}, or both \citep{2011A&A...529A.118T}.

We have searched for new RCB stars among the entire ASAS-3 south survey dataset using both light variations and IR excess techniques to increase our overall detection efficiency. The survey has monitored millions of stars with a declination lower than +28\degr, since October 2000. The survey characteristics are described in section~\ref{sec_obs}. RCBs are mainly known to undergo unpredictable, fast, and large photometric declines (up to 9 mag over a few weeks) due to carbon clouds forming close to the line of sight that obscure the photosphere. First, we searched for such signatures among the ASAS-3 light curves of pre-selected objects presenting a near-IR excess using the 2MASS magnitudes. We applied this classical searching method to the ASAS-3 South ACVS1.1 variable stars catalogue and then to the entire ASAS-3 south source catalogue. Second, we visually inspected the light curves of all objects selected by \citet{2012A&A...539A..51T,2012TissInPrepa} for their warm shells. The details of these analyses are discribed in section~\ref{sec_ana} and a comparison of their results is discussed in section~\ref{sec_discu}. The spectroscopic confirmation of 23 new RCB stars is presented in section~\ref{sec_spectro}, and each star is discussed individually in section~\ref{sec_stars}. A discussion about their maximum brightness, spatial distribution and spectral energy distribution can be found respectively in sections ~\ref{sec_maxmag}, \ref{sec_spdistrib} and \ref{sec_SED}.

We note also that we searched for DY Per stars, which might be the cooler counterpart of RCB stars. We found two new DY Per stars among the entire ASAS-3 south survey dataset. More details on DY Per stars and the analysis can be found in section~\ref{ana_dyper}.

\section{Observational data \label{sec_obs}}

The All Sky Automated Survey (ASAS) \citep{1997AcA....47..467P} has been monitoring the entire southern sky and part of the northern sky ($\mathrm{\delta}< +28\degr$) since October 2000, using large field of view CCD cameras and a wide-band V filter. Sources brighter than V$\sim$14 mag were catalogued (i.e. $\sim 10^7$ objects). The system is located at Las Campanas Observatory. The light curve of each individual object can be downloaded from the ASAS-3 website: http://www.astrouw.edu.pl/asas/. The CCD resolution is about 14.8\arcsec/pixel and therefore the astrometric accuracy is around 3-5\arcsec~for bright stars, but could be  up to 15\arcsec~for fainter ones. With such resolution the photometry in crowded fields is uncertain. 

Spectroscopic follow-up of 104 RCB candidates, 24 known RCB stars and 6 known HdC stars was performed with the Wide Field Spectrograph (WiFeS) instrument \citep{2007Ap&SS.310..255D} attached to the 2.3 m telescope at Siding Spring Observatory of the Australian National University. WiFeS is an integral field spectrograph permanently mounted at the Nasmyth A focus. It provides a $25\arcsec \times38\arcsec$~field with 0.5 arcsec sampling along each of twenty-five $38\arcsec \times1\arcsec$~slitlets. The visible wavelength interval is divided by a dichroic at around 600 nm feeding two essentially similar spectrographs. Observations are presented with a 2-pixel resolution of 2 \AA{}.  The spectra were obtained during seven observational runs:  27-29 November 2009, 14-17 July 2010, 9-13 December 2010,  20-25 July 2011, 4-7 June 2012, 23-25 July 2012 and 1-3 August 2012. They are presented in figures \ref{sp_ASASnew}, \ref{sp_RCBKnown}, \ref{sp_IRAS_V391Sct}, \ref{sp_DYPers}, \ref{sp_3hotRCBs} and \ref{sp_HdC}.

Additional spectra were obtained with the Reticon detector and Unit spectrograph on the 1.9m telescope at the Sutherland site of the South African Astronomical Observatory (SAAO). All spectra were obtained with grating 6 which has a resolution of $\sim$2.5 \AA{}, and a useful range of about 3600-5400~\AA{} at the angle setting used. The spectra were taken between 1989 and 1994. They have been extracted, flat-field corrected, sky-subtracted and wavelength calibrated. A flux calibration was done for the spectra using the spectrophotometric standard LTT7379 although due to the slit and variable seeing, the data are not photometric.  Eight spectra were obtained of V532 Oph. These were combined by taking a median of all the spectra. All spectra are plotted in figure \ref{sp_Kilkenny}.

\begin{figure*}
\centering
\includegraphics[scale=0.55]{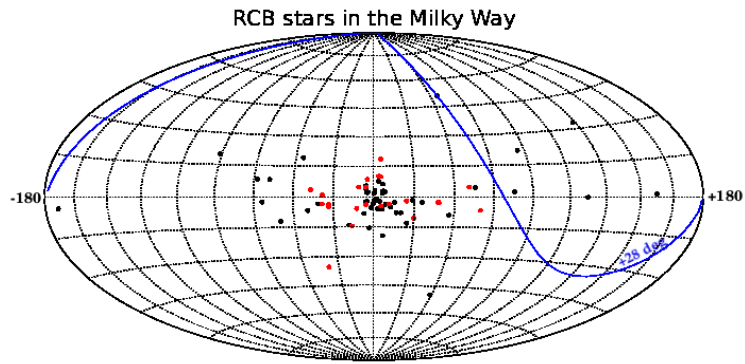}
\caption{Galactic distribution (l,b) of RCB stars in the Milky Way centered at (0,0). The black dots are RCB stars known before this analysis, the red ones are newly discovered. The blue line represents the limit of observation of the ASAS-3 south survey with a declination of +28\degr.}
\label{fig_distrib}
\end{figure*}

\section{Analysis \label{sec_ana}}

Many large scale surveys are actually underway and new RCB stars can be found among the millions of light curves produced, although the aperiodic large brightness variations of RCB stars make them difficult to find and a cautious approach is necessary. A significant RCB star sample also needs to be gathered and characterised before more automatic searches can be made as the light curves are so diverse. Furthermore, the ASAS-3 dataset is affected by systematic problems that result in some artificial declines which may be misleading, as well as the annual gap in the observations, and variations of the survey depth.

For our initial approach, we decided to apply a series of selection criteria to keep interesting candidates for spectroscopic follow-up. The selection criteria were defined to optimise the number of new RCBs found but also to minimise the quantity of light curves for visual inspection. First, as the vast majority of RCBs possess a warm circumstellar dust shell, we kept only objects that present an IR excess using the 2MASS J, H and K magnitudes (see Figure~\ref{IRfig}). Second, we selected objects that show large amplitude brightness variations and/or show a rapid decline rate in luminosity. These criteria differ slightly depending on the ASAS-3 catalogues being analysed. They are described in sections~\ref{sec_searchRCB_classic_varcat} and \ref{sec_searchRCB_classic_allcat}.

This classical analysis was performed in two steps, named analyses A and B in Table~\ref{tab.NewRCB}. Initially, we focused on the ASAS-3 variable stars catalogue, ACVS1.1. We then decided to direct our efforts towards the entire ASAS-3 source star catalogue, as we found that some previously discovered RCB stars were not listed in ACVS1.1. The details of both analyses are described below. 

The limiting magnitude of the ASAS-3 survey is about $\mathrm{V_{lim}}\sim 14$ mag. We estimated that we can detect RCB stars with a maximum magnitude no fainter than $\mathrm{V_{RCB,max}}\sim 13$ mag, which corresponds to maximum distances of $\sim$40 and $\sim$20 kpc, respectively for RCBs with absolute magnitudes $\mathrm{M_V}$ of -5 and -3.5 mag \citep{2009A&A...501..985T}. We also expected that most RCBs located in the Galactic Bulge and/or close to the Galactic plane, where the density of known RCBs is highest \citep{2008A&A...481..673T}, would not be in the ASAS-3 catalogue for two reasons:  first, RCBs would be fainter due to high interstellar extinction; second, the photometry and source detection become complicated with the high density of stars and the low CCD resolution of $\sim$15\arcsec/pixel. Of the 19 RCBs already known towards the Galactic Bulge \citep{2005AJ....130.2293Z,2008A&A...481..673T,2011A&A...529A.118T}, only 2 (EROS2-CG-RCB-3 and OGLE-GC-RCB-2) were bright enough to be catalogued and monitored by the ASAS-3 survey, each with a maximum magnitude of $\mathrm{V_{max}}\sim 13.6$ mag. The former star has only ten measurements on its light curve, while the latter has more than 800. However, we note that the decline observed in the OGLE-III dataset \citep[Figure 5]{2011A&A...529A.118T}, around HJD~2453500 days, is not clearly detectable in the ASAS-3 light curve of OGLE-GC-RCB-2 due to the limiting magnitude of the survey, and therefore this RCB star would not have been found despite its large number of measurements.

To increase our detection efficiency and compare the pros and cons of different methods we decided also to visually inspect the ASAS-3 light curves of all objects listed in \citet{2012A&A...539A..51T}, where no selection criteria based on the light curve were used, but only 2MASS \citep{2006AJ....131.1163S} and WISE \citep{2010AJ....140.1868W} IR colours to select objects with hot, bright circumstellar shells.  The related analysis is discussed in section~\ref{ana_wise} and labelled C and D in Table~\ref{tab.NewRCB}.

\subsection{Search for RCB stars in the ACVS1.1 catalogue (Analysis A) \label{sec_searchRCB_classic_varcat}}

This catalogue is a concatenation of five different catalogues of variable stars as detected by the ASAS collaboration: first with declinations $< 0\degr$ and various ranges of right ascension, $0^{h}-6^{h}$\citep{2002AcA....52..397P}, $6^{h}-12^{h}$ \citep{2003AcA....53..341P}, $12^{h}-18^{h}$\citep{2004AcA....54..153P}, $18^{h}-24^{h}$ \citep{2005AcA....55...97P}, and with declinations $<+28\degr$ \citep{2005AcA....55..275P}. ACVS1.1 is available on the ASAS Web page\footnote{ASAS Web page: http://www.astrouw.edu.pl/asas/?page=catalogues} and contains 50124 entries. For each object listed such parameters of variability as periodicity, amplitude of variation $\mathrm{V_{amp}}$ and maximum brightness, $\mathrm{V_{max}}$, are given, as well as the IRAS and 2MASS photometry.

We used 2MASS colours and the $\mathrm{V_{amp}}$ parameter to select stars for visual inspection. We selected all objects that present an IR excess as described in Figure \ref{IRfig}, but added a supplementary constraint on the amplitude of variation of $\mathrm{V_{amp}}>1$ mag for objects with $\mathrm{H-K}<0.2$. A total of 1685 objects were selected. After visual inspection of their light curves to search for typical RCB declines in brightness, 17 stars were selected for spectroscopic follow-up. The rejected objects were mostly periodic variable stars, such as Miras and eclipsing binary systems.

Seven stars presented unmistakable RCB-type declines in brightness and, indeed, their respective spectra confirmed their true nature as RCB stars. They are listed in Table \ref{tab.NewRCB} and named ASAS-RCB-1 to -7. Another star, named ASAS-RCB-8, has a light curve with a relatively slow decline, but its spectrum shows that it is a warm RCB star. These new RCB stars are discussed individually in section~\ref{sec_stars}. Their spectra and light curves are presented in Figures~\ref{sp_ASASnew} and \ref{fig_lc1} to \ref{fig_lc4} respectively.

Three of these RCBs, ASAS-RCB-3, -4 and -6, were also recently reported and spectroscopically confirmed by \citet{2012ApJ...755...98M}. They analysed the light curves of each object listed in the ACVS1.1 catalogue using a machine-learned algorithm to classify objects and applied a hard cut on the probability that an object belongs to the RCB class. They followed up their candidates spectroscopically and found four RCB stars, the three that have just been mentioned plus ASAS-RCB-9. We did not find ASAS-RCB-9 in Analysis A because the 2MASS IR magnitudes given in the ACVS1.1 catalogue are not correct, certainly due to a crossmatching issue. \citet{2012ApJ...755...98M} technique is a very interesting method to discover rare objects such as RCB stars, but it would benefit from more training with a larger sample of known RCB stars' light curves. The newly found RCB stars reported here could be used to improve the algorithm and increase its detection efficiency. As we discuss below, other RCB stars can still be found in the ASAS-3 dataset and the \citet{2012ApJ...755...98M} algorithm could be of great use in finding RCB stars that are at the limit of detectability.

Surprisingly, despite our success in revealing eight new RCB stars, we did not rediscover all 29 previously known bright RCB stars that could have potentially been listed in the ACVS1.1 catalogue. Indeed, 11 of the 29 known RCB stars were not recovered. This is easily explained for four of the stars: Y Mus, DY Cen, V739 Sgr and MV Sgr, as they did not show any brightness variation in their respective ASAS-3 light curves. However, this is not the case for 7 other known RCBs, namely SU Tau, UX Ant, UW Cen, V348 Sgr, GU Sgr, RY Sgr and V532 Oph. They were not found even though they exhibited large amplitude declines or recovery phases on their overall ASAS-3 light curves and had significant IR excesses.  In fact, those 7 RCBs are not even in the ACVS1.1 catalogue. This is due to the fact that the ACVS1.1 catalogue contains only stars where variability occurred during the first few years of ASAS-3 data, and therefore stars with large brightness variations later in the survey have been missed. Table~\ref{tab.KnownRCB} illustrates this by showing the difference between the maximum amplitude listed in the ACVS1.1 catalogue and the one taking into account the entire ASAS-3 light curves. This outcome indicates that there may still be a number of undiscovered RCB stars in the entire ASAS-3 south source star catalogue. The analysis of this catalogue is described below.

\begin{figure}
\centering
\includegraphics[scale=0.4]{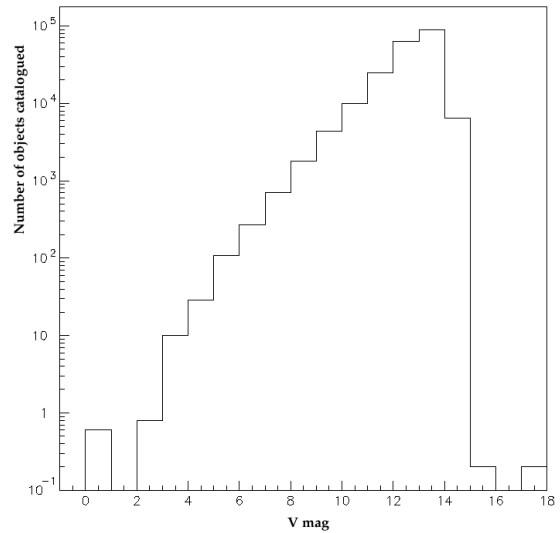}
\caption{Distribution of the V magnitude of objects listed in the entire ASAS-3 south source catalogue used in this study. }
\label{distrib_mag}
\end{figure}

\subsection{Search for RCB stars in the entire ASAS-3 south source star catalogue (Analysis B) \label{sec_searchRCB_classic_allcat}}

About 12 million stars were catalogued and monitored by the ASAS-3 south survey. Thanks to the ASAS collaboration, we obtained the source star catalogue, which contains the coordinates of each object, as well as a V magnitude truncated to its integer value, calculated from the 30 brightest measurements. This catalogue contains all objects located south of the celestial equator with more than 200 measurements, and northern objects, up to +28\degr, with at least 100 measurements. These constraints were applied to assure the validity of each object. We treated each entry individually, and duplicates were removed during the final stages of the analysis.

\begin{figure}
\centering
\includegraphics[scale=0.5]{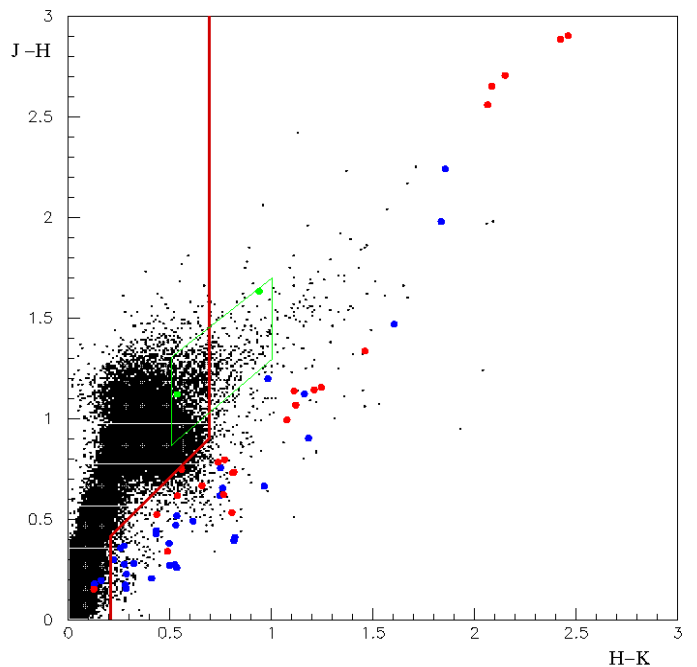}
\caption{$\mathrm{J-H}$ vs $\mathrm{H-K}$ diagrams using the 2MASS magnitudes of all objects listed in the ASAS-3 ACVS1.1 variable stars catalogue (black dots). The red lines show the selection criteria used in Analysis A. The blue points correspond to the RCB stars that were previously known, the red points to the 23 newly discovered RCB stars. The green lines represent the selection box used to find new DY Per stars.  The green points correspond to the 2 new DY Per stars. }
\label{IRfig}
\end{figure}

First, we kept only stars brighter than $V<13$ mag as we found it difficult to recognise RCB photometric declines for fainter objects, knowing that the ASAS-3 limiting magnitude is $\mathrm{V_{lim}}\sim 14$ mag. The brightness distribution of all sources catalogued is presented in Figure~\ref{distrib_mag}. After the magnitude cut, a total of 6038078 objects remained. Second, we matched each object with the 2MASS catalogue using a progressive matching radius technique, from 5\arcsec  to 15\arcsec. Only 2541 ASAS-3 objects did not have an association with the 2MASS catalogue. Third, we applied the same selection criteria that were applied to the ACVS1.1 catalogue, selecting all objects with an IR excess, except that this time we added a strict cut at $\mathrm{H-K}>0.2$ (see Figure~\ref{IRfig}). About 67000 objects passed this last selection and their respective ASAS-3 light curves, spanning 10 years (2000-2010), were downloaded through the ASAS-3 Web interface\footnote{ASAS Web interface: http://www.astrouw.edu.pl/asas/?page=aasc} using an automatic script. Using magnitudes calculated from the smallest photometric aperture, we calculated for each object the total amplitude of variation $\mathrm{V_{amp}}$, based on four measurements at both maximum and minimum brightness, and the maximum rate of decline in brightness, $\mathrm{D_{max}}$, based on five consecutive measurements. We were careful to reject light curves with systematic photometric issues that can mimic a large decline. These were recognisable as they occurred on many light curves during the same night.  The number of fast declines occurring per night was therefore counted over all light curves and data from nights presenting 10 or more fast declines were rejected.  Finally, we visually inspected 5264 objects with $\mathrm{V_{amp}}>1.5$ mag and an additional 1710 objects with $\mathrm{D_{max}}>0.02$ mag.day$^{-1}$. From these candidates, 60 objects were chosen for spectroscopic follow-up. This method enabled us to confirm 4 new RCB stars, named ASAS-RCB-9 to -12. They are listed in Table~\ref{tab.NewRCB} and their spectra and light curves are presented in Figures~\ref{sp_ASASnew}  and \ref{fig_lc2} respectively.

Of the 28 known RCB stars located in the sky area monitored by the ASAS-3 south survey (strictly below dec$<+28\degr$, R CrB is therefore not counted), we succeeded this time in rediscovering 23 of them with this technique. As mentioned in the previous analysis, V739 Sgr, DY Cen, MV Sgr and Y Mus do not present any photometric declines and could not therefore have been found. (We note that Y Mus would not have been detected regardless as it presents only a weak IR excess.) The last known RCB star not rediscovered is V348 Sgr simply because it is not listed in the ASAS-3 south source catalogue that we used, despite the fact that its light curve is available from the ASAS Web site and presents multiple declines. It has not been listed because only 95 measurements were counted, which is lower than the threshold of 200 used to form the source catalogue.
 
Among the three strong RCB candidates suggested by \citet{2012A&A...539A..51T}, only two were selected by the analyses that we have just described: ASAS-RCB-7 with Analysis A and ASAS-RCB-9 with Analysis B. The remaining suggested candidate is V581 CrA. Its ASAS-3 light curve is available from the interactive ASAS Web service with more than 200 measurements. It was not found by both analysis, despite the fact that it has gone through large photometric variations (see Figure~\ref{fig_lc3}) and its 2MASS colours indicate an IR excess that would have been selected by our criteria. We obtained a spectrum and confirmed its membership of the RCB class of stars. We named it ASAS-RCB-13.  We did not find an entry for that star, nor in the ACVS1.1 catalogue, nor in the entire ASAS-3 south source catalogue. Many of its photometric measurements are fainter than V$\sim$14 mag, and therefore, not all of them passed our quality requierements. The total number of qualified measurements dropped below the threshold to select ASAS-RCB-13 in the ASAS-3 source catalogue.

The fact that V348 Sgr and ASAS-RCB-13 are missing indicated to us that other RCB stars may yet be revealed from the ASAS-3 survey. Despite the success of our method of finding new RCB stars, we may also have missed some because of the 2MASS association method we used to comply with ASAS-3 astrometric uncertainty (up to 15 arcseconds), or because of the strict limit applied on the amplitude and rate of decline,  or the limit on the number of measurements used to validate the objects that were subsequently analysed. Finally, we note that some RCBs may also have remained faint during the period covered and could have recovered their brightness after 2010, and some RCBs may have been in a faint phase on the reference images used to create the ASAS-3 source catalogue. 

\subsection{Search for RCB stars using the \citet{2012A&A...539A..51T,2012TissInPrepa} catalogue (Analyses C and D)\label{ana_wise}}

The analyses described above have a limiting magnitude of $V\sim13$, but with a strong decrease in detection efficiency for objects fainter than $V\sim12$. To complete our search of the ASAS-3 south catalogue, we decided to download and visually inspect the ASAS-3 light curves of each object listed in the catalogue of RCB candidate stars that was created from the 2MASS survey and the WISE preliminary data release \citep{2012A&A...539A..51T}, and expanded using the WISE final all-sky data release \citep{2012TissInPrepa}. The selection was made in two steps. First, all WISE selected objects were cross-matched with the entire ASAS-3 source catalogue and the light curves of the matching candidates were visually inspected (Analysis C). Second, in case some RCBs were not included in the ASAS-3 source catalogue (as was found to be the case with ASAS-RCB-13), we interrogated the ASAS-3 database to download the light curves of all remaining WISE objects selected (Analysis D). Overall, 22 objects were selected for spectroscopic follow-up, but 4 were too faint during our observational runs\footnote{Namely ASAS J143647-6404.5, J173401-2951.8, J174317-1824.0 and J183631-2059.2. The last star was classified as an M type star and is therefore most likely a Mira variable star \citep{1967IBVS..228....1H}.}. Of the 18 remaining, 8 show clear features due to the C$_2$ and CN molecules and weak abundance of $^{13}$C at 4744 and 6260 $\AA{}$. They are named ASAS-RCB-14 to -21 and are listed in Table ~\ref{tab.NewRCB}. Their spectra are presented in Figure~\ref{sp_ASASnew} and light curves in Figures~\ref{fig_lc3} and \ref{fig_lc4}.

None of these 8 new RCB stars would have been selected by the classical methods described in sections~\ref{sec_searchRCB_classic_varcat} and \ref{sec_searchRCB_classic_allcat}, for three reasons. First, the maximum total amplitude variation observed for all of them is lower than the selection criterion applied of 1.5 mag. Second, five present a maximum luminosity fainter than $V\sim13$. Finally,  the light curves of five of them, namely ASAS-RCB-14, -15, -19, -20 and -21, possess too few measurements to be selected as valid objects in the entire ASAS-3 source catalogue we analysed. Indeed, these obects only have between 10 and 50 measurements. More photometric follow-up will be needed for these RCBs in order to detect more aperiodic fast declines and to definitely confirm their nature.

Of the 12 new RCBs that were discovered by classical methods (ASAS-RCB-1 to -12), only two were not found by our new method: ASAS-RCB-1 and -8, the two warmest RCBs of the series. It is also worth emphasizing that ASAS-RCB-13 and -21 were found only because they were listed in the \citet{2012A&A...539A..51T} catalogue. None of them was listed in the ASAS-3 south source catalogue we used, demonstrating the difficulty of searching for RCB stars in that dataset.

Clearly the release of the post-2010 data of the ASAS-3 south survey and of course those from the ASAS-3 north survey will be the source of new RCB discoveries. Dedicated algorithms such as the one developed by \citet{2012ApJ...755...98M} will be very useful, especially if they combine their search with near- and mid-IR broadband photometry.

\subsection{Search for DY Per stars in the entire ASAS-3 south source star catalogue \label{ana_dyper}}

Only two DY Per stars were known in our Galaxy: the prototype DY Persei \citep{2007A&A...472..247Z,2009arXiv0905.4344Y} and EROS2-CG-RCB-2 \citep{2008A&A...481..673T}. Recently, \citet{2012ApJ...755...98M} have reported four possible new ones and \citet{2012MNRAS.424.2468S} discussed in details the reasons why two bright cool N-type carbon stars, V1983 Cyg and V2074 Cyg, should also be ascribed to this type of stars. DY Per stars are known to be cooler than typical RCB stars (with $\mathrm{V-I}\sim2$ mag, $\mathrm{T_{eff}}\sim3000-3500$ K). Unlike RCBs which typically have sharp declines and slow recoveries, DY Per stars decline symmetrically with the fading and recovery happening at similar rates. Some examples of these brightness variations can be found in \citet{2001ApJ...554..298A} and \citet{2004A&A...424..245T,2009A&A...501..985T}, where 13 Magellanic DY Per stars were reported. \citet{2009AcA....59..335S} have also presented a list of Magellanic DY Per candidates with their respective OGLE-III light curves. Many of them are interesting and will need spectroscopic follow-up for confirmation. Since so few DY Per stars are known, the characterisation of this class of stars is not complete. DY Per stars are known to be cool carbon-rich supergiant stars, with a possible hydrogen deficiency  \citep{1997PASP..109..969K,2007A&A...472..247Z} and an isotopic $^{13}$C/$^{12}$C ratio higher than RCBs. They are surrounded by circumstellar dust shells and they form new dust as seen in light curve variations. It is shown in \citet{2009A&A...501..985T} that there exists a relation between the maximum depth of the optical brightness variation and the brightness of the shell in mid-IR.

DY Per stars might be the cooler counterpart of RCB stars, and there may therefore be an evolutionary connection between both classes of stars.
A DY Per star's circumstellar dust shell is on average hotter ($\mathrm{T_{eff}}\sim800-1500$ K) than an RCB's shell \citep{2009A&A...501..985T}. It is interesting to note that a sharp peak was observed at $\sim$11.3 $\mu$m in the mid-IR spectrum of DY Persei itself, obtained by \citet{2011ApJ...739...37A} with the Spitzer Infrared Spectrograph, indicating the possible presence of polycyclic aromatic hydrocarbons (PAHs) in its shell, which are not observed in the majority of RCB stars\footnote{Two RCB stars, one hot and one warm, DY Cen and V854 Cen, show PAH features in their respective spectra \citep{2011ApJ...739...37A,2011ApJ...729..126G}. These are also the two RCB stars with the highest measured hydrogen abundance\citep{1998A&A...332..651A}.}. The presence of PAHs in DY Persei needs to be confirmed with spectroscopic observation at wavelengths shorter than 10 $\mu$m. 

From the DY Per characteristics that have just been mentioned, it is clear that DY Per stars are also very similar to ordinary carbon stars with an N type spectrum and further studies will be needed to investigate if they are indeed two different types of object.

We searched for new DY Per stars in the ASAS-3 south dataset by inspecting visually a sample of objects selected in a $\mathrm{J-H}$ versus $\mathrm{H-K}$ 2MASS colour diagram (see the selected area in Figure~\ref{IRfig}). DY Per stars are known to look similar to ordinary carbon stars in broadband photometry. For objects with $0.3\leqslant \mathrm{H-K} \leqslant0.5$, we added a supplementary constraint on the amplitude of variation, $\mathrm{V_{amp}}>1.5$ mag. A total of 12421 light curves were visually inspected to search for the typical DY Per symmetric decline and 10 candidates were kept for further photometric follow-up. Of these 10 candidates, we confirmed 2 new DY Per  stars. Their spectra show C$_2$ and CN features and clear $^{13}$C enhancement at $\lambda\sim4744 \AA{}$. They are named ASAS-DYPer-1 and -2 and are listed in Table~\ref{tab.NewRCB}. ASAS-DYPer-1 is even brighter by at least 1.2 mag in K than the prototype DY Persei (see Table~\ref{tab.IR}). Their respective light curves and spectra are presented in Figures ~\ref{fig_lc5} and \ref{sp_DYPers}. We present the classification of the other 8 candidates in Table~\ref{tab.Otherstars}: 5 show a spectrum with TiO features similar to Mira type stars, 2 show nebula-like emission lines, and the eighth is a hot supergiant star with very narrow hydrogen lines.

\section{Spectroscopic confirmation \label{sec_spectro}}

RCB stars are of spectral type K to F with an effective temperature ranging mostly between 4000 and 8000 K \citep{2012A&A...539A..51T}, with a few exceptions being hotter than 10000 K (V348 Sgr, DY Cen,  MV Sgr and HV 2671) \citep{2002AJ....123.3387D}. The standard RCB atmospheric composition can be found in \citet[Table 1]{1997A&A...318..521A}. The main characteristics of the spectrum are hydrogen deficiency, with H/He ranging mostly between $10^{-6}$ and $10^{-4}$ (V854 Cen is the exception with H/He$\sim10^{-2}$); carbon richness equivalent to C/He$\sim1\%$; and N enrichment, with C/N$\sim$1.3 \citep{2000A&A...353..287A}. A complete summary and discussion of the usual abundances found in RCB stars in the context of hot and cold WD merger scenarios can be found in \citet{2011MNRAS.414.3599J}.

HdC stars are also hydrogen-deficient and carbon-rich stars, but there are no photometric declines on record. Only 6 HdC stars are currently known and their effective temperature range is between 5500 and 7000 K \citep{2012A&A...539A..51T}. Mid-infrared data indicate that there is no sign of the existence of a circumstellar shell and therefore they do not produce carbon dust clouds as RCB stars do. The exception is HD 175893 which has a circumstellar dust shell ($T\mathrm{_{eff}}\sim 500$ K) but has never been seen to undergo a decline phase \citep[Figure 12]{2012A&A...539A..51T}. HD 175893 could therefore play an important role in the understanding of an evolutionary link between RCB and HdC stars.

The optical spectra of 24 known RCB stars, 3 hot RCBs and the 6 known HdC stars are presented respectively in Figures \ref{sp_RCBKnown} and \ref{sp_Kilkenny}, \ref{sp_3hotRCBs} and \ref{sp_HdC}. They were obtained by the SSO/2.3m/WiFeS and the SAAO/1.9m/Reticon/Unit spectrographs which cover two different wavelength ranges, respectively 3400 to 9700 \AA{} and 3600 to 5400 \AA{}. A summary and further details of these spectra can be found in Section~\ref{sec_obs} and Table~\ref{tab.KnownRCB}.

The optical spectrum of an RCB star depends on its temperature and also on its brightness phase. At maximum brightness, a classical RCB star will show easily identifiable features due to the C$_2$ and CN molecules \citep{1983MNRAS.202P..31B} and no sign of hydrogen (H$\mathrm{\alpha}$ or CH),  while a hot RCB star presents a spectrum with many emission lines (Figure~\ref{sp_3hotRCBs}). During a decline phase, carbon features are still recognisable but a few emission lines start to emerge, with the NaI (D) lines being the more prominent one (see the spectra of ES Aql, ASAS-RCB-13 and ASAS-RCB-19 for examples). A detailed discussion of RCB spectra during a decline phase can be found in \citet{1996PASP..108..225C}. The absence of enhanced $^{13}$C is also a characteristic of RCB stars \citep{1994MNRAS.268..544P}, even if few RCB stars do have detectable $^{13}$C: V CrA, V854 Cen, VZ Sgr, and UX Ant have measured $^{12}$C/$^{13}$C $<$25 \citep{1989MNRAS.238P...1K,2008MNRAS.384..477R,2012ApJ...747..102H}. We attempted to estimate the abundance of $^{13}$C using the $^{13}$C$^{12}$C band-head at 4744 \AA{}, and also the $^{13}$CN band at 6260 \AA{} \citep{1991MNRAS.249..409L}. For all new RCB stars with clear carbon features in their spectrum, it was possible to determine the low content of $^{13}$C in their atmosphere. As most of the new RCB stars found are located in the Galactic bulge, this study was made difficult by the strong interstellar reddening that limits the amount of flux in the blue part of the spectrum.

Overall, we systematically followed up spectroscopically 104 stars: 23 of which were revealed to be RCB stars and 2 others to be DY Per stars. Their spectra are presented in Figures \ref{sp_ASASnew}, \ref{sp_IRAS_V391Sct} and \ref{sp_DYPers}. The classification of the remaining 81 objects is given in Tables~\ref{tab.Otherstars} and \ref{tab.Otherstars2}. The two main families of rejected objects are the Mira variables and T Tauri stars.  Miras present large photometric variations and a spectrum rich in oxygen. We classified them using the M giant spectra library provided by \citet{1991AJ....101..662B}. T Tauri stars present hydrogen-rich spectra with emission lines over a large range of temperatures (9 rejected objects are classified as Orion type objects in the SIMBAD database). Nine other stars presenting carbon features were not considered as RCB stars. They were rejected because they presented clear $\mathrm{H\alpha}$ emission in three cases and/or high abundance of $^{13}$C for the 6 others. The latter group passed the selection criteria for spectroscopic follow-up because the IR selection criterion, $\mathrm{H-K}>0.7$ mag, was intentionally defined so as not to be too restrictive. They are located in the continuation of the classical giant branch evolutionary track in the $\mathrm{J-H}$ vs $\mathrm{H-K}$ diagram inside the selection criteria area of DY Per stars shown in Figure~\ref{IRfig}. Their light curves show variations up to 2 mag, but with no clear signs of a fast decline. As they all present large photometric oscillations of $\sim$0.8 mag amplitude and their spectra do not show clear signs of the presence of hydrogen, they should be considered as DY Per star candidates. 

Stars that were considered as RCB candidates in the literature were also followed up \citep{1996PASP..108..225C,1997IAUC.6632....1H,2012JAVSO..40..539C}. We confirmed IRAS 1813.5-2419 and V391 Sct, but rejected five other stars: BL Cir and GM Ser are Miras,  with the spectral type M6 and M7 respectively; V1773 Oph is an F star;  V1860 Sgr is a giant G6III star; and V2331 Sgr is an S star that presents weak TiO and CN features, but no ZrO band-head.

Among the 23 new RCB stars, 4 present a spectrum with many metallic absorption lines, particularly C I, indicating that their effective temperature is relatively high, $\mathrm{T_{eff}}> 7500$ K. They are ASAS-RCB -8, -10, -12 and V391 Sct. For the other new RCBs, one can use the triplet absorption lines due to ionised calcium ($\lambda\sim$ 8498, 8543 and 8662 \AA{}). Indeed, \citet{1971ApJ...167..521R} shows that the intensity of these lines is a good indicator of  carbon stars temperature; the cooler the temperature, the weaker the lines. This could be an important tool as most of the spectra are strongly reddened, but it should be used with caution as we are not certain that this observation is applicable to the warmer and specific carbon stars that are RCBs. In the spectra presented in Figures~\ref{sp_ASASnew} and \ref{sp_IRAS_V391Sct}, one can separate the 19 remaining RCB stars into 3 groups of increasing temperature based on the strength of the Ca-II triplet absorption lines: \textit{Weak}: ASAS-RCB-4, -5, -13, -14, -19 ; \textit{Moderate}: ASAS-RCB-2, -6, -7, -15, -16, -17, -18 and -21 ; \textit{Strong}: ASAS-RCB-1, -3, -9, -11, -20 and IRAS 18135-2419. We also note that the 2 DY Per stars do not present any Ca-II triplet absorption lines, indicating that they are indeed cool carbon stars (see Figure \ref{sp_DYPers}).

\section{Maximum magnitude \label{sec_maxmag}}

The median of the maximum brightness found for the newly detected RCB stars is $\mathrm{V_{max,med}}\sim 12.5$. That is only 1.5 magnitude brighter than the magnitude limit of the ASAS-3 south survey. These magnitudes are listed in Table~\ref{tab.NewRCBvar}. We cannot be certain that the real maximum optical brightness was reached during the ASAS-3 survey for 9 RCBs as it lasted less than 20 days. They are ASAS-RCB-2, -7, -12, -13, -14, -15, -19, -20 and -21. Five of these (ASAS-RCB-14. -15, -19, -20 and -21) have very few measurements in their light curves. A further group of five RCBs (ASAS-RCB-2, -12, -15, -19 and -20) present a high IR excess, $\mathrm{J-K}>4.5$ (see table~\ref{tab.IR} and figure~\ref{IRfig}), which indicates that they are in a phase of high dust production activity.

\begin{figure*}
\centering
\includegraphics[scale=0.35]{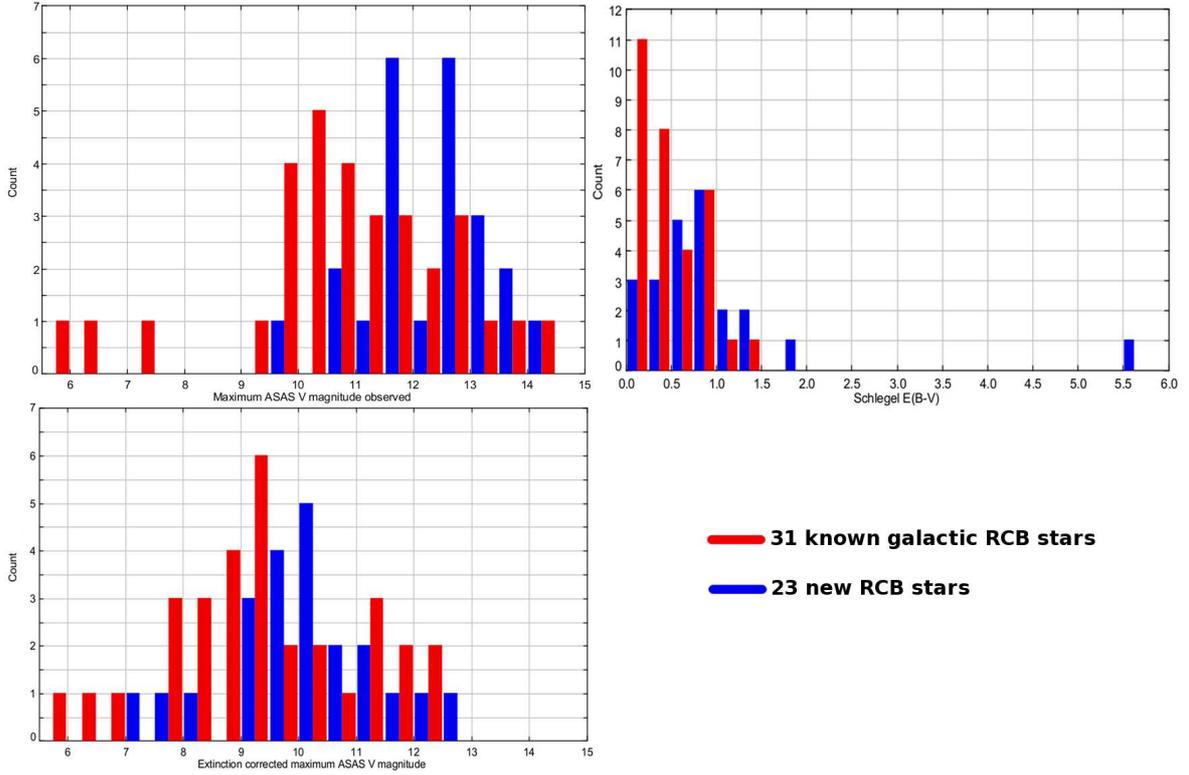}
\caption{Top-left: Distribution of the maximum ASAS-3 V magnitude observed for the known Galactic RCB stars (red, see Table~\ref{tab.KnownRCB}) and the 23 new RCB stars (blue, see Table~\ref{tab.NewRCBvar}). Bottom-left: Same distribution but corrected for interstellar extinction using the \citet{1998ApJ...500..525S} reddening map and $\mathrm{R_V} = 3.1$. (ASAS-RCB-9 was not plotted here as we considered its reddening value to be overestimated and not valid.) Top-right: Reddening $\mathrm{E(B-V)}$ for the two lists of objects.}
\label{fig_maxmag}
\end{figure*}

A comparison of the maximum ASAS-3 V magnitudes of the 31 known Galactic RCB stars listed in Table ~\ref{tab.KnownRCB} and the 23 new RCBs is shown in Figure~\ref{fig_maxmag}, before and after correcting for interstellar extinction. The interstellar extinction $\mathrm{A_V}$ was calculated using the \citet{1998ApJ...500..525S} $\mathrm{E(B-V)}$ reddening map with 4-pixels interpolation and a coefficient $\mathrm{R_V} = 3.1$. Interestingly, the apparent maximum brightness of the new RCBs is fainter on average than the already known RCBs. After correction for interstellar extinction, however, this difference is strongly reduced as the new RCBs are affected on average by a stronger interstellar reddening (see Figure~\ref{fig_maxmag}, top-right). Overall the Galactic RCBs' maximum magnitude peaks at $\mathrm{V_{ASAS}}\sim 9.8$ mag which is within the range expected for RCB stars located at an average distance of $\mathrm{D_{GC}}\sim 8.2$ kpc\footnote{It corresponds to a  distance modulus of 14.6 mag.}, the distance of the Galactic centre \citep{2012arXiv1208.1263N}. Indeed, with an absolute magnitude ranging between $-5\leqslant \mathrm{M_V}\leqslant-3.5$ \citep[see][]{2009A&A...501..985T}, RCBs located at that distance would have an apparent V magnitude between 9.6 and 11.1 mag. 

\section{Distribution of Galactic RCB stars  \label{sec_spdistrib}}

The small number of known RCB stars and the biases affecting their discovery in different surveys have prevented us from obtaining a clear picture of their true spatial distribution.  \citet{1985ApJS...58..661I}  reported a scale height of $h\sim400$ pc assuming $\mathrm{M_{bol}} = -5$ and concluded that RCBs are part of an old disk-like distribution. However, \citet{1998PASA...15..179C} noted that the Hipparcos velocity dispersion of RCB stars is similar to that of other cool hydrogen-deficient carbon stars and extreme helium stars, suggesting that RCB stars might have a bulge-like distribution. \citet{1986ASSL..128....9D} and \citet{2005AJ....130.2293Z} came to a similar conclusion. Recently, \citet{2008A&A...481..673T} also found that the majority of Galactic RCB stars seem to be concentrated in the bulge but with the surprising peculiarity of being distributed in a thin disk structure ($61<\mathrm{h^{RCB}_{bulge}}<246$ pc, 95\% c.l.).  In the Large Magellanic Cloud, RCBs are distributed mostly along the bar \citep[Figure 2]{2009A&A...501..985T}.

The growing number of known RCBs allows us to form a clearer picture of their Galactic distribution. Indeed, their maximum apparent magnitude (Section~\ref{sec_maxmag}) and their spatial distribution (Figure~\ref{fig_distrib}) indicate that Galactic RCB stars do in fact a have bulge-like distribution.

\section{Spectral energy distribution (SED)\label{sec_SED}}

\begin{figure*}
\centering
\includegraphics[scale=0.9]{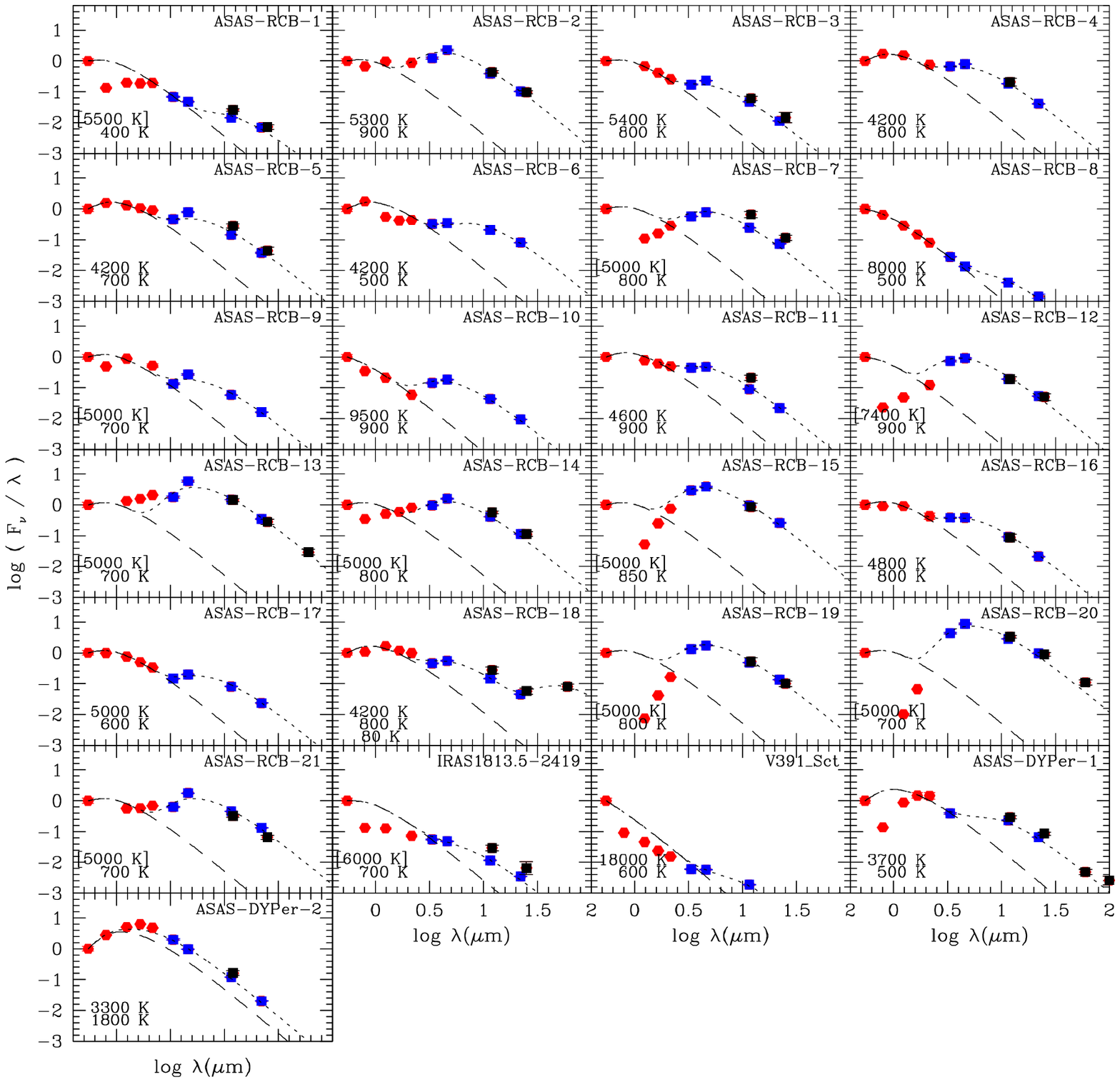}
\caption{Best attempts to reconstruct the spectral energy distribution of the 23 new RCB stars and the 2 new DY Pers. This task was made difficult by high interstellar extinction and by the fact that the series of optical and infrared measurements were not taken at the same time and therefore not at the same brightness phase (decline or maximum?). See section~\ref{sec_SED} for more details. The red dots represent the extinction-corrected optical and infrared measurements: the maximum V magnitude as observed on the ASAS light curve (Table~\ref{tab.NewRCBvar}), the DENIS I measurement and the infrared 2MASS J, H and K  or the DENIS J and K magnitudes. The blue and black dots correspond respectively to the WISE and IRAS measurements. The blackbody of the photosphere is represented with dashed lines, while the dotted lines represent the best fit found with one or two circumstellar shells added. The temperatures of these simple blackbodies are listed for each star (top: photosphere, bottom: shell). If the photospheric temperature is indicated between square brackets, the related blackbody was added only for representation as there are no convincing measurements to constrain the fit. }
\label{fig_SED}
\end{figure*}

\begin{figure*}
\centering
\includegraphics[scale=0.55,angle=270]{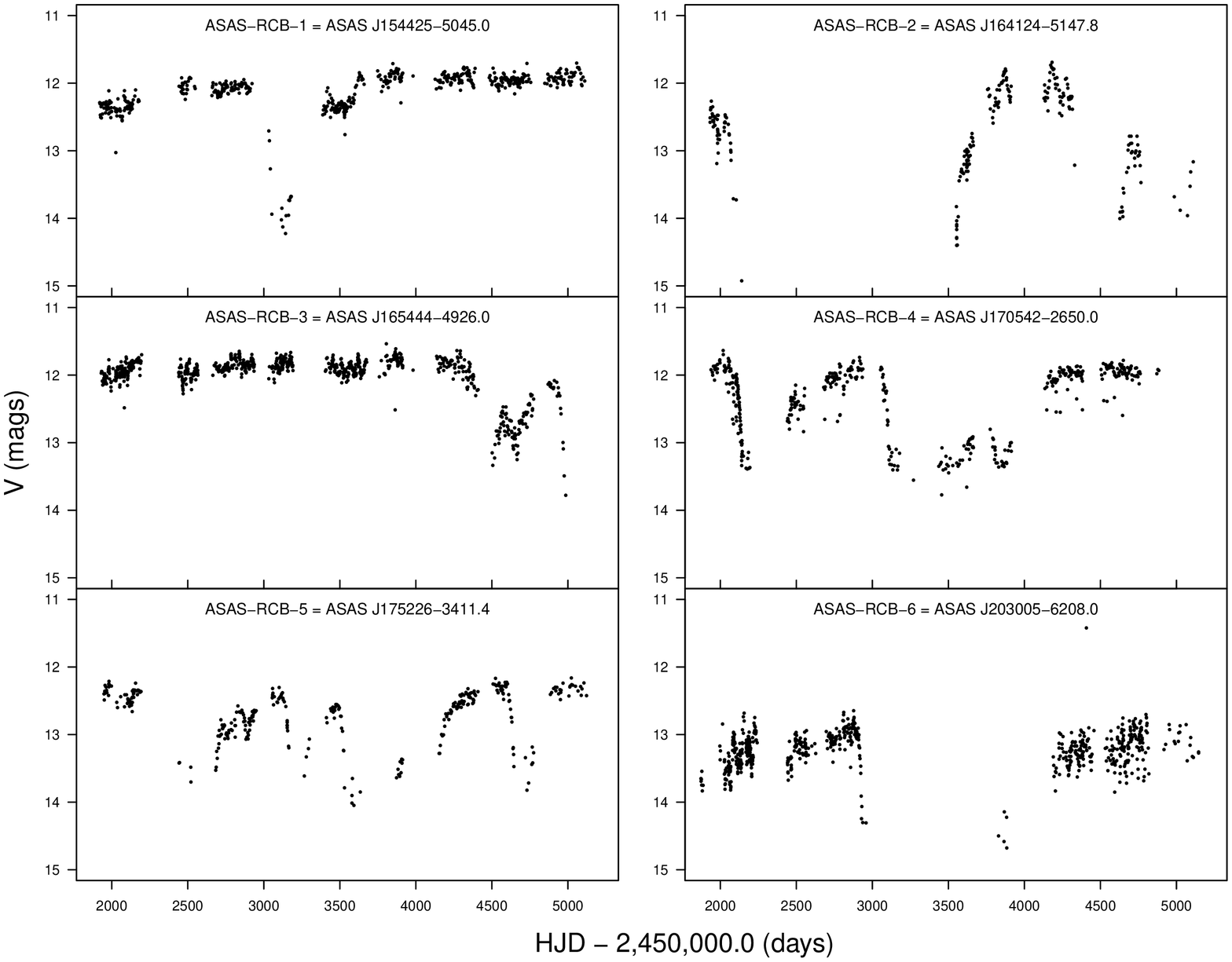}
\caption{ASAS-3 light curves of ASAS-RCB-1 to -6 (ASAS measurement grade A, B or C). }
\label{fig_lc1}
\end{figure*}

\begin{figure*}
\centering
\includegraphics[scale=0.55,angle=270]{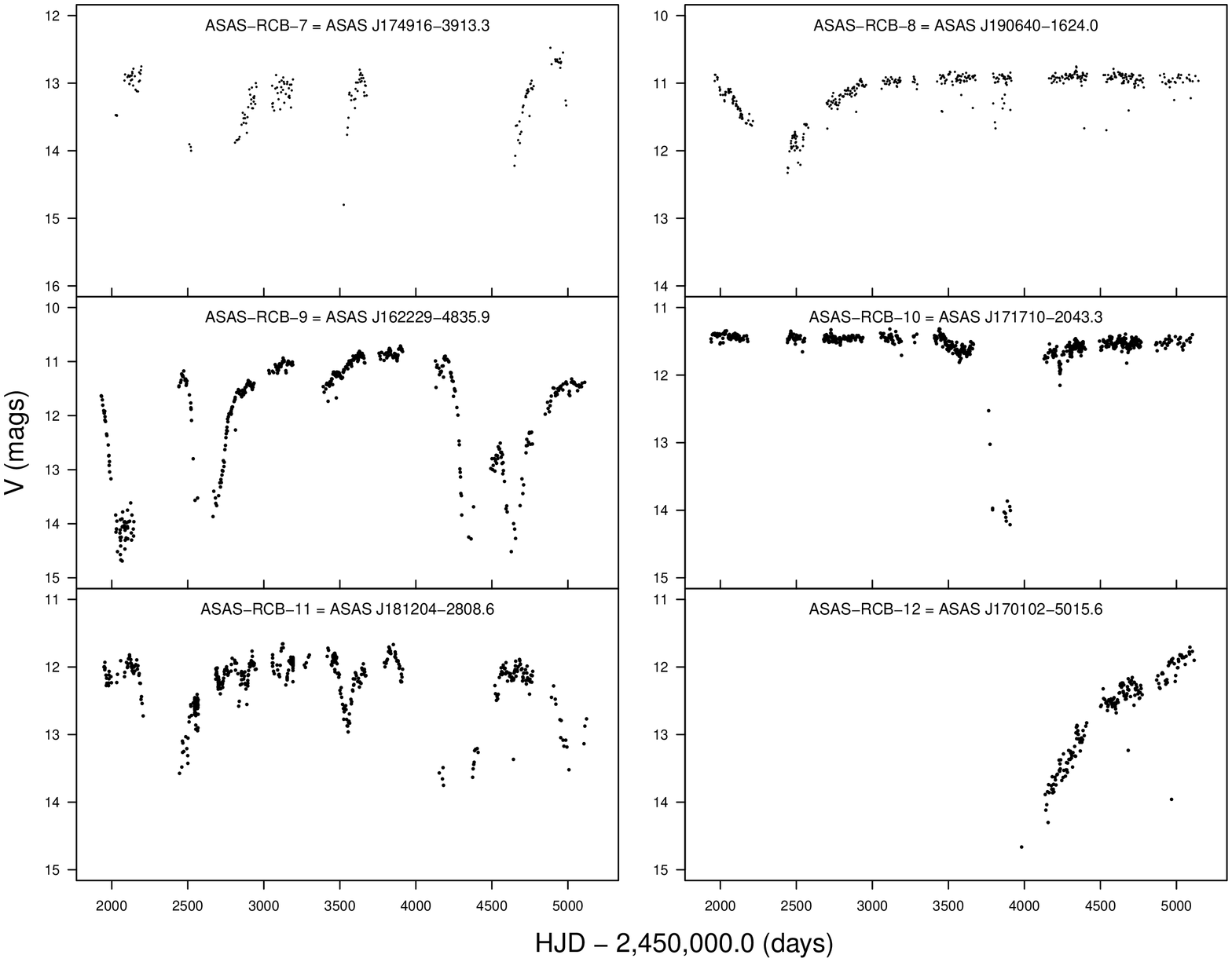}
\caption{ASAS-3 light curves of ASAS-RCB-7 to -12 (ASAS measurement grade A, B or C).}
\label{fig_lc2}
\end{figure*}

\begin{figure*}
\centering
\includegraphics[scale=0.55,angle=270]{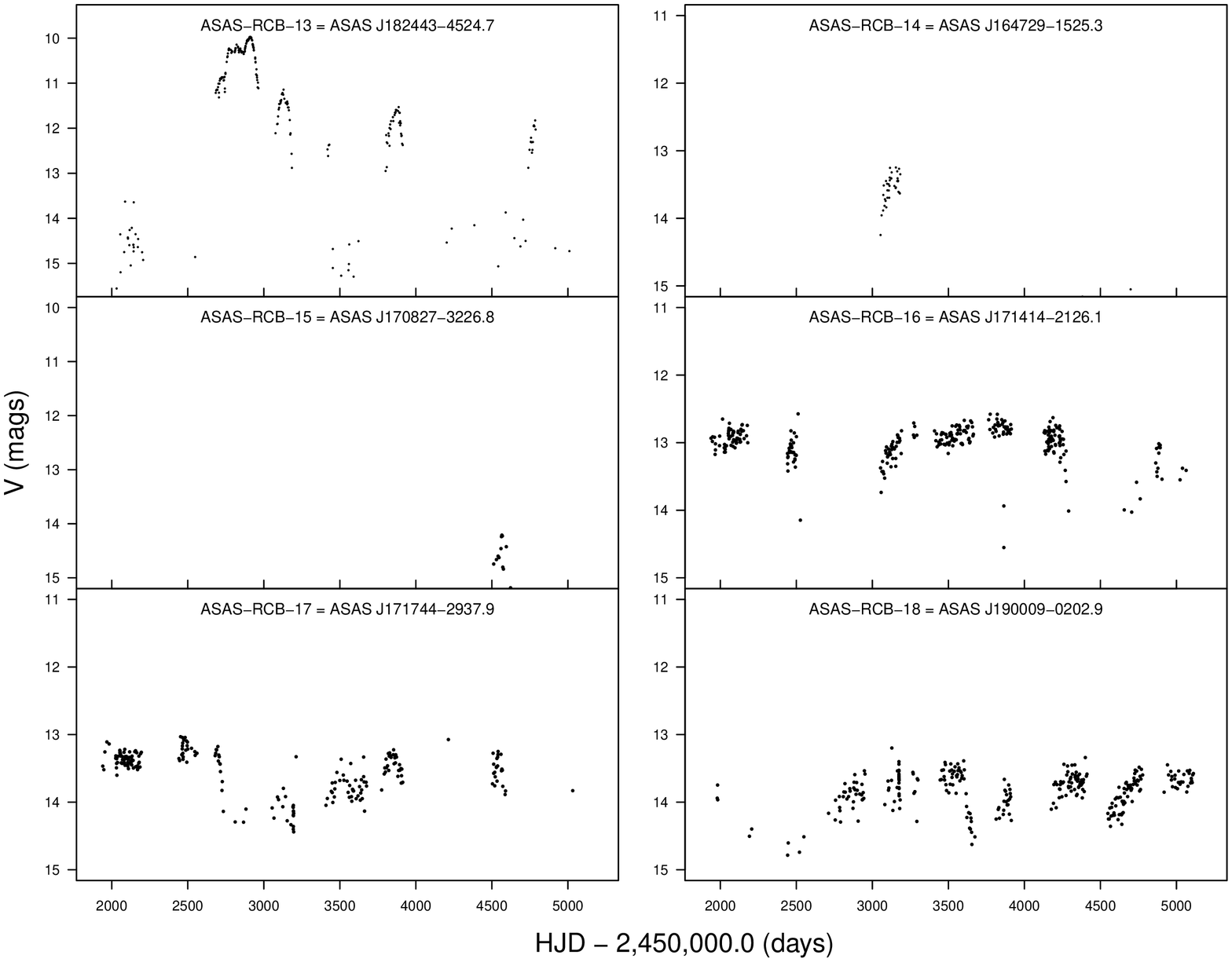}
\caption{ASAS-3 light curves of ASAS-RCB-13 to -18 (ASAS measurement grade A, B or C).}
\label{fig_lc3}
\end{figure*}

\begin{figure*}
\centering
\includegraphics[scale=0.55,angle=270]{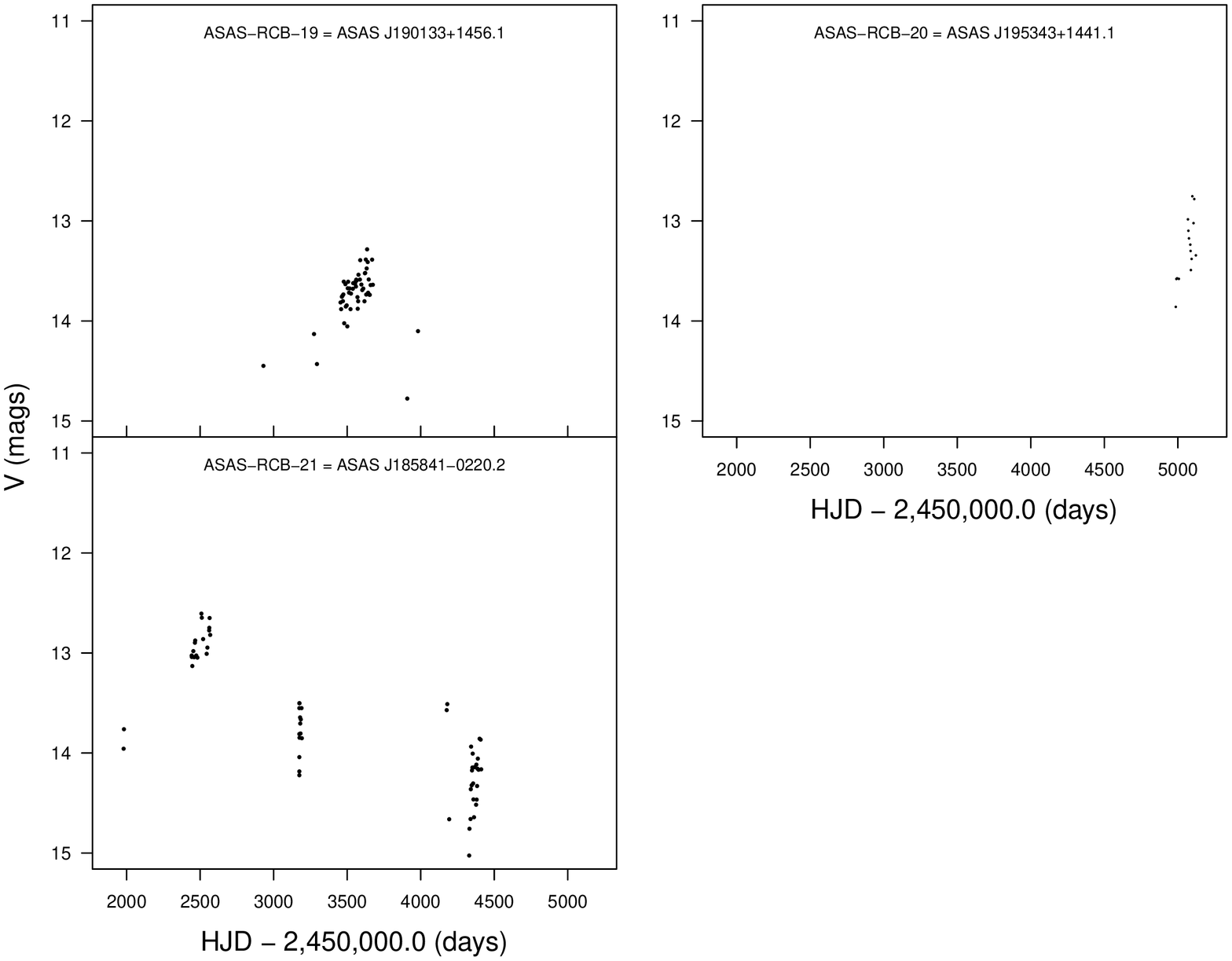}
\caption{ASAS-3 light curves of ASAS-RCB-19 to -21 (ASAS measurement grade A, B or C).}
\label{fig_lc4}
\end{figure*}

\begin{figure*}
\centering
\includegraphics[scale=0.55,angle=270]{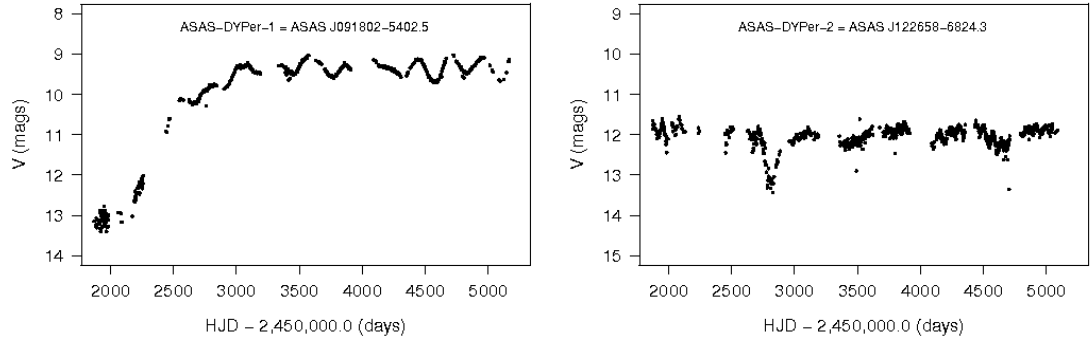}
\caption{ASAS-3 light curves of ASAS-DYPer-1 and -2 (ASAS measurement grade A, B or C).}
\label{fig_lc5}
\end{figure*}

\begin{figure*}
\centering
\includegraphics[scale=1.0]{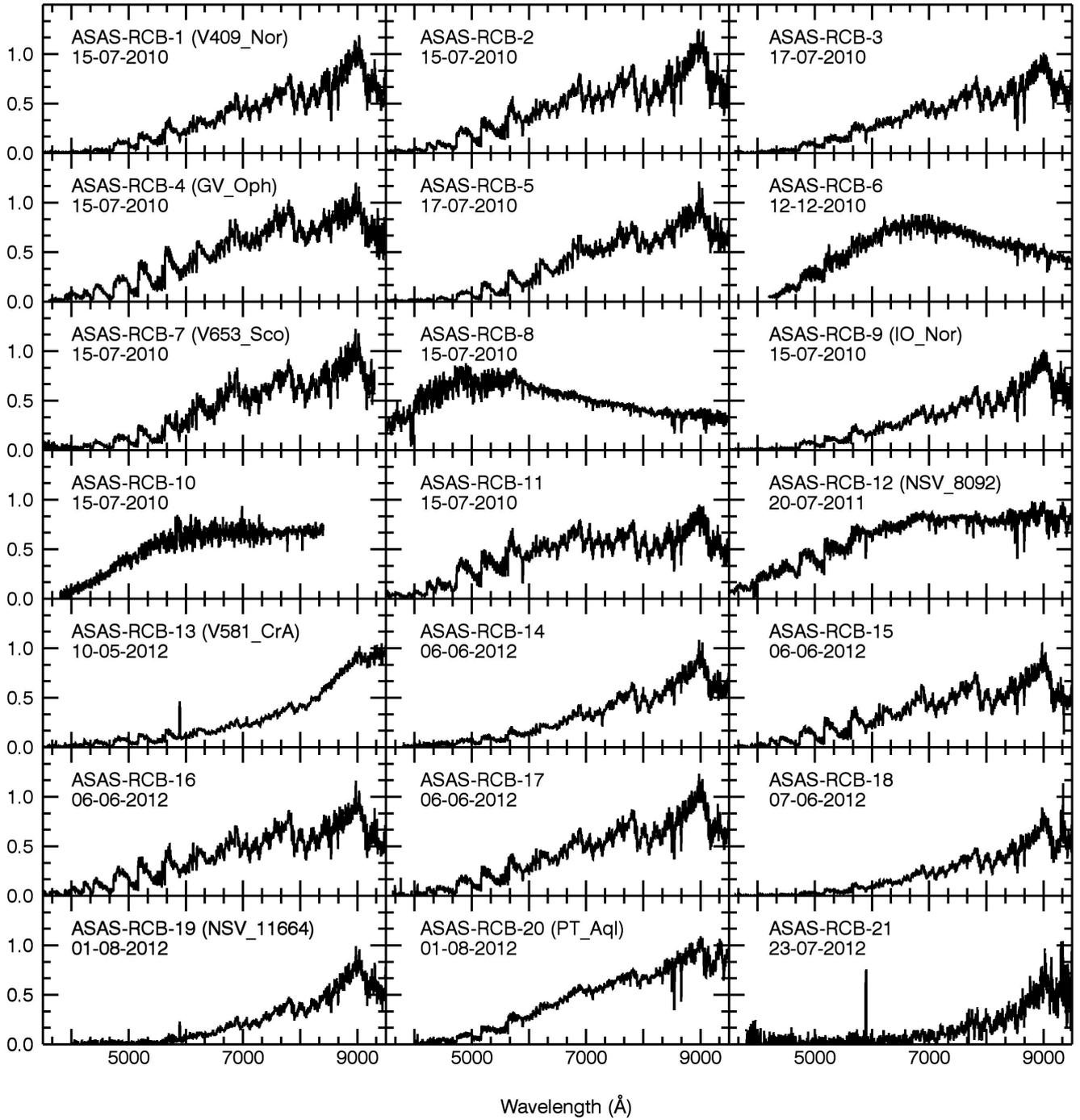}
\caption{Spectra of the 21 new RCB stars found by the analyses of the ASAS-3 catalogues, using the SSO/2.3m/WiFeS spectrograph.}
\label{sp_ASASnew}
\end{figure*}

\begin{figure*}
\centering
\includegraphics[scale=1.0]{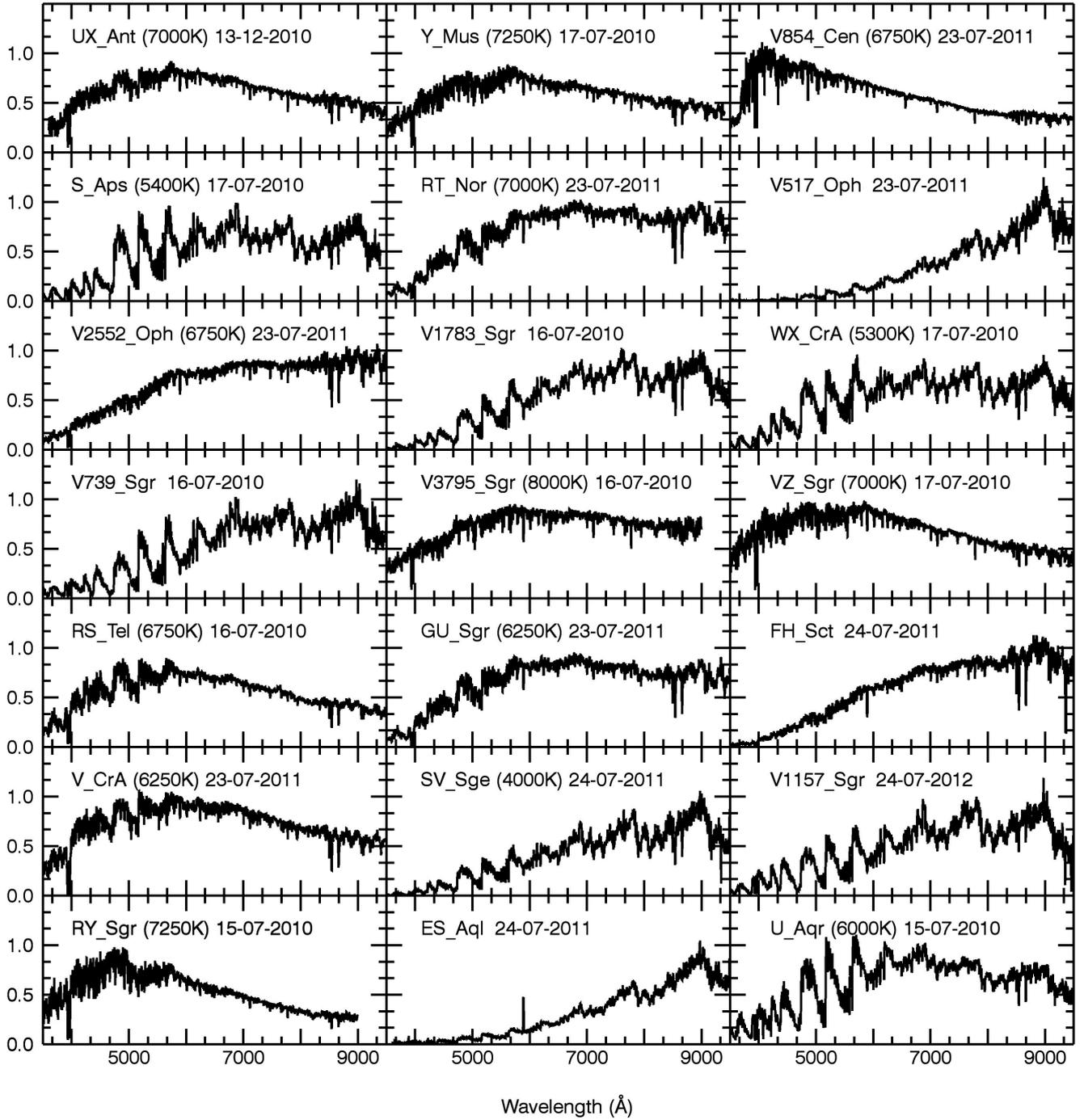}
\caption{Spectra of 21 already known RCB stars observed with the SSO/2.3m/WiFeS spectrograph.}
\label{sp_RCBKnown}
\end{figure*}

\begin{figure*}
\centering
\includegraphics[scale=1.0]{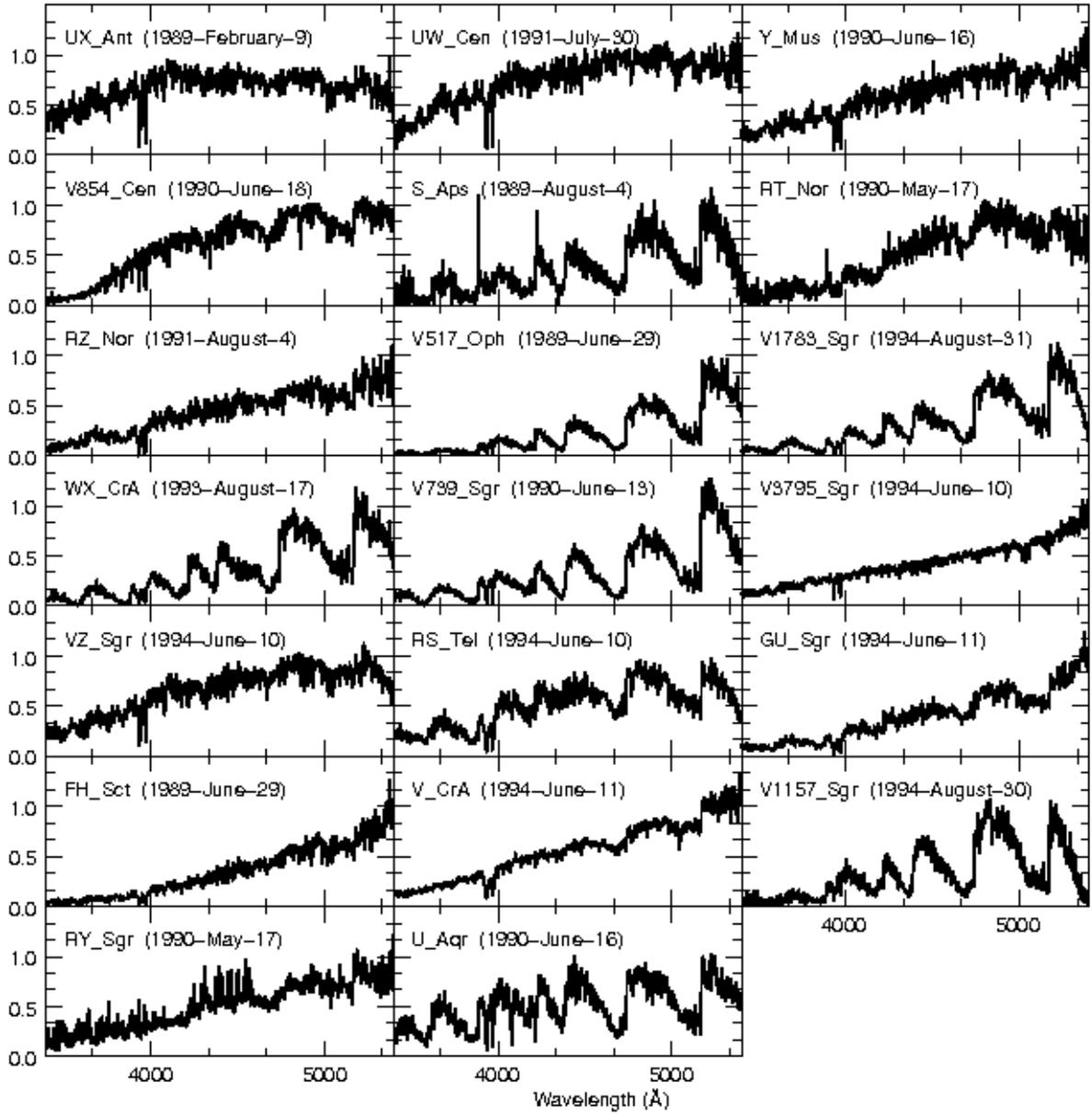}
\caption{Spectra of 20 already known RCB stars from 3600 to 5400 \AA{} using the SAAO/1.9m/Reticon/Unit spectrograph.}
\label{sp_Kilkenny}
\end{figure*}

\begin{figure*}
\centering
\includegraphics[scale=1.2]{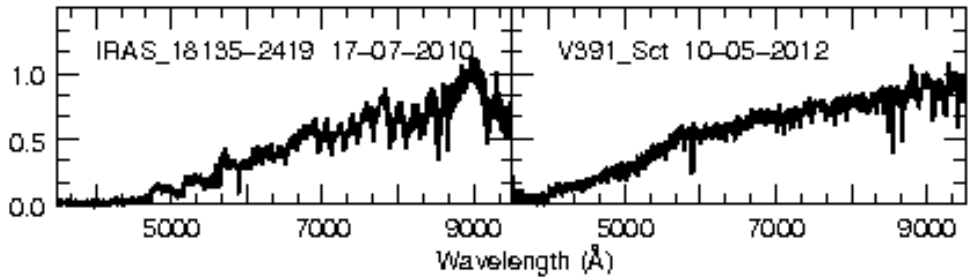}
\caption{Spectra of IRAS 1813.5-2419 and V391 Sct  observed with the SSO/2.3m/WiFeS spectrograph.}
\label{sp_IRAS_V391Sct}
\end{figure*}

\begin{figure*}
\centering
\includegraphics[scale=1.2]{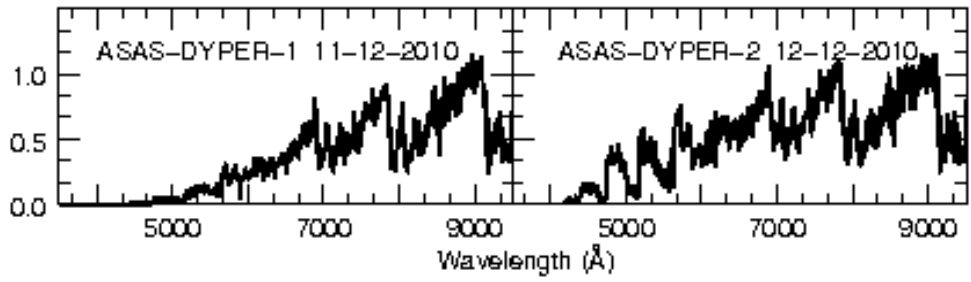}
\caption{Spectra of ASAS-DYPer-1 and -2 observed with the SSO/2.3m/WiFeS spectrograph.}
\label{sp_DYPers}
\end{figure*}

\begin{figure*}
\centering
\includegraphics[scale=1.1]{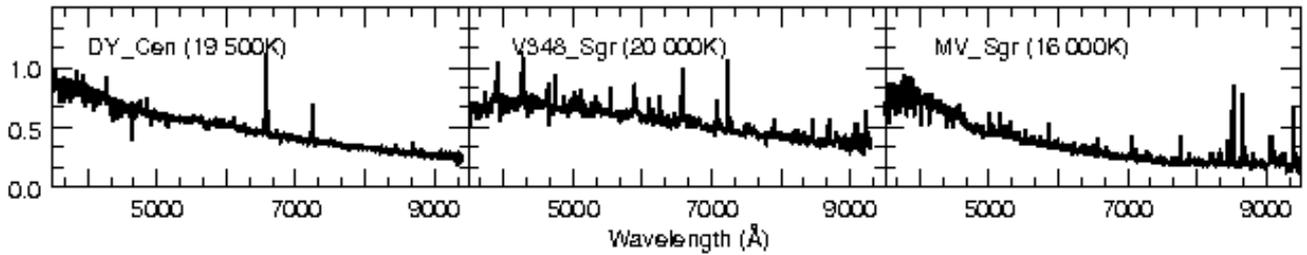}
\caption{Spectra of three hot known RCB stars observed with the SSO/2.3m/WiFeS spectrograph.}
\label{sp_3hotRCBs}
\end{figure*}

\begin{figure*}
\centering
\includegraphics[scale=1.0]{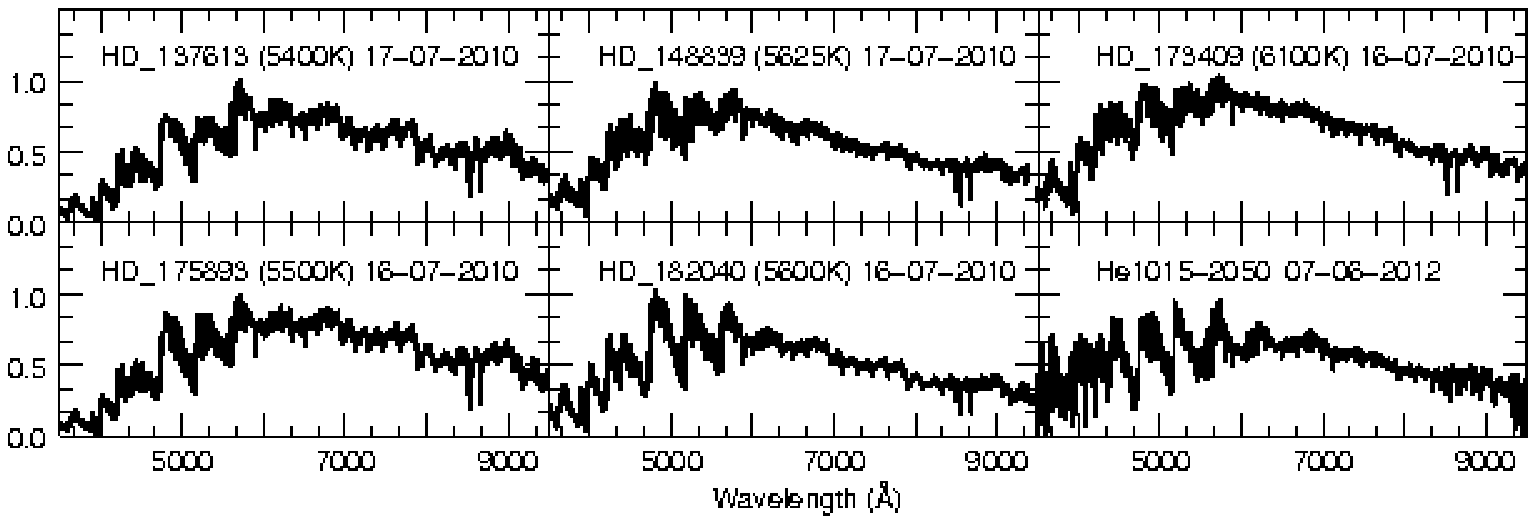}
\caption{Spectra of the six known  HdC stars observed with the SSO/2.3m/WiFeS spectrograph.}
\label{sp_HdC}
\end{figure*}

We attempted to estimate the photosphere temperature for the newly found RCBs by fitting simple blackbodies to the series of optical, infrared and mid-infrared measurements collected by the following surveys: DENIS \citep{1994Ap&SS.217....3E}, 2MASS \citep{2006AJ....131.1163S}, WISE \citep{2010AJ....140.1868W} and IRAS \citep{1988iras....7.....H}. The observed maximum ASAS V magnitude and the IR DENIS and 2MASS magnitudes that we used are listed respectively in Tables~\ref{tab.NewRCBvar} and ~\ref{tab.IR}. RCB star photospheres are known to have a temperature between 4000 and 8000 K, with a few hot RCBs having a temperature higher than 10000 K, while DYPer photospheres have a temperature between 3000 and 3500 K. The spectral energy distribution (SED) of 34 known RCBs are presented in \citet{2012A&A...539A..51T} and some have had their effective temperature determined from spectral synthesis analyses applied to high resolution spectra \citep[see][]{2000A&A...353..287A}. RCBs are warmer than classical carbon-rich stars, which have a temperature ranging mostly between 2000 and 3000 K. Classical carbon-rich stars' SED are described in detail by \citet{2001A&A...369..178B}. The reconstructed SED of the newly discovered RCBs are presented in Figure~\ref{fig_SED}. Three major issues prevented us from obtaining strong constraints on photosphere temperatures:  the unsynchronised measurements,  high interstellar extinction and the unknown real maximum brightness for some of the RCBs.

Indeed, except for ASAS-RCB-2, -10 and -16, the DENIS and 2MASS epochs (listed in Table~\ref{tab.IR}) were all taken before the start of the ASAS-3 survey. We do not therefore know the RCBs' luminosity phase (decline or maximum?) at the time of these IR observations. This demonstrates the necessity of long-term monitoring. Nevertheless, we attempted to overcome this limitation by using either the DENIS J and H or the 2MASS J, H and K magnitudes to construct the SEDs according to four rules:
\begin{enumerate}
\item If  $\mathrm{\delta J} = \mathrm{J_{2MASS}} -  \mathrm{J_{DENIS}} > 0.2$ mag: the epoch during the 2MASS measurements was considered to have been taken under a higher dust extinction phase than the DENIS measurement. Only the three uncorrected DENIS magnitudes (I, J and K) were used.
\item If  $\mathrm{\delta J} =\mathrm{ J_{2MASS}} - \mathrm{ J_{DENIS}} < -0.2$ mag: the epoch during the DENIS measurements was considered to have been taken under a higher dust extinction phase than the 2MASS measurement. The 2MASS J, H and K magnitudes were used as well as the DENIS I magnitude corrected by a carbon extinction law \citep{1995A&A...293..463G}. That is $\mathrm{I^{corr}_{DENIS}} = \mathrm{I_{DENIS}} - \mathrm{A^{carbon}_I}$, with $\mathrm{A^{carbon}_I} = 1.66 \times \mathrm{A^{carbon}_J} = 1.66 \times \mathrm{\delta J} $.
\item If $\mathrm{J_{2MASS}}$ and $\mathrm{J_{DENIS}}$ agrees within $\pm$0.2 mag, only the 2MASS J, H and K magnitudes were used as well as the uncorrected DENIS I magnitude.
\item If no DENIS measurements exist, only the 2MASS J, H and K magnitudes were used.
\end{enumerate}

From Figure~\ref{fig_SED}, it appears clearly that the DENIS and 2MASS measurements of, respectively, ASAS-RCB-12 and ASAS-RCB-1, -7, -15, -19, -20,  IRAS 1813.5-2419 and ASAS-DYPer-1 were obtained during a decline phase of luminonisity, when dust clouds obscured the photosphere. We cannot correct for this effect and therefore did not use these magnitudes to fit the SED.

The interstellar extinction correction was applied for all optical and infrared measurements following the reddening $\mathrm{E(B-V)}$ values listed in Table~\ref{tab.NewRCBvar}.  The relative extinction coefficients $\mathrm{A_I}/\mathrm{A_V}$,  $\mathrm{A_J}/\mathrm{A_V}$, $\mathrm{A_H}/\mathrm{A_V}$ and  $\mathrm{A_K}/\mathrm{A_V}$ determined by \citet[Table 3]{1989ApJ...345..245C} were used. Their values are respectively 0.479, 0.282, 0.19 and 0.114. The $E(B-V)$ reddening value listed for ASAS-RCB-9 is overestimated and not valid. We fitted it to be $\sim$1.0 mag by assuming the photosphere temperature to be 5000 K.

As already discussed in section~\ref{sec_maxmag}, we note that it is uncertain whether or not the real maximum optical brightness was reached during the ASAS-3 survey for 9 RCBs. Therefore, the maximum ASAS V magnitudes used in the SED of ASAS-RCB-2, -7, -12, -13, -14, -15, -19, -20 and -21 should be considered as a lower limit.

The results are shown in Figure~\ref{fig_SED}. Our best attempt to fit two blackbodies over the SED using only convincing measurements is also presented. Overall, the circumstellar shell temperatures are well constrained, thanks to the WISE and/or IRAS surveys. Evidence of shell brightness variation is also visible, especially for ASAS-RCB-7, -11 and IRAS 1813.5-2419. The errors on the shell fitted temperatures are in the order of 50 K.  The photosphere temperature depends strongly on the extinction correction applied and on whether or not the measurements were really obtained during a maximum brightness phase. As their respective spectra show, ASAS-RCB-8, -10 an V391 Sct were found to be hot RCBs ($\mathrm{T_{eff}}>7500$ K).  Six RCBs, i.e. ASAS-RCB-4, -5, -6, -11, -16 and -18,  appear to have a relatively cool photosphere, with $4000<\mathrm{T_{eff}}<5000$ K, while another three, i.e. ASAS-RCB-2, -3 and -17, seem warmer with $5000<\mathrm{T_{eff}}<6000$ K. Also, as expected, the temperature of both DYPer stars lies between 3000 and 4000 K. The photosphere temperatures for these 14 stars are considered to be a fair estimate, but we underline that they are only indicative and should be used with caution. We did not succeed in determining a consistent temperature estimate for the eleven remaining RCB stars, i.e. ASAS-RCB-1, -7, -9, -12, -13, -14, -15, -19, -20, -21 and IRAS 1813.5-2419.

Interestingly, we found that ASAS-RCB-18 was detected  by the IRAS survey at $\sim$65 $\mu$m. This measurement suggests that there exists a second circumstellar shell of cooler temperature, $\sim$80$\pm$30 K, which may be the remnant of an older evolutionnary phase.

\begin{table*}
\caption{Known Galactic RCB stars: ASAS-3 id and variability as observed in the ASAS-3 survey between December 2000 and October 2009
\label{tab.KnownRCB}}
\medskip
\centering
\begin{tabular}{lccccccc}
\hline
\hline
Name & ASAS-3 id & Number of  & \multicolumn{2}{c}{Maximum amplitude (mag)} & Maximum & \multicolumn{2}{c}{Spectrum} \\
&  & declines & ASAS-3  & ACVS1.1  & brightness &  \multicolumn{2}{c}{date} \\
&& observed &   (2000-2009)  & ($\mathrm{V_{amp}}$) & $\mathrm{V_{max}}$ (mag)  & SSO/WiFeS & SAAO/Reticon\\
\hline
\multicolumn{8}{c}{Galactic}\\ 
\hline
\object{SU Tau} & J054904+1904.3 & 3 & 4.5 & \_ & 9.8 & & \\
\object{UX Ant} & J105709-3723.9 & 3 & 2.8 & \_ & 12.0 & 13-12-2010 & 09-02-1989\\
\object{UW Cen} & J124317-5431.7 & 2 & 4.2 & \_ & 9.3 & & 30-07-1991\\
\object{Y Mus}  & J130548-6530.8  & none & \_ & \_ & 10.25 & 17-07-2010 & 16-06-1990\\
\object{DY Cen} & J132534-5414.7  &  none  & \_ &  \_ & 13.2 & 17-07-2010 & \\
\object{V854 Cen} & J143450-3933.5  & 9 & 3.9 & 3.91 & 7.0 & 23-07-2011 & 18-06-1990\\
\object{S Aps}  & J150924-7203.8  & 3 & 4.5 & 4.61 & 9.7 & 17-07-2010 & 04-08-1989\\
\object{R CrB}  & J154834+2809.4  &  1 & 7.1 & 7.51 & 5.8 & & \\
\object{RT Nor}  & J162419-5920.6  & 1 & 2.0 & 0.19 & 10.1 & 23-07-2011 & 17-05-1990\\
\object{RZ Nor}  & J163242-5315.6  & 4 & 3.2 & 0.63 & 10.3 & $^a$ & 04-08-1991\\
\object{V517 Oph} & J171520-2905.6  & 8 & 3.0 & 1.30 & 11.4 & 23-07-2011 & 29-06-1989\\
\object{V2552 Oph} & J172315-2252.0 & 3 & 3.1 & 2.23 & 10.9 & 17-07-2010 &\\
 & & & & & & 23-07-2011 &\\
\object{V532 Oph}  & J173243-2151.6 & 3 & 3.0 & \_ & 11.8 & & \\
\object{OGLE-GC-RCB-2} & J175357-2931.7 & 1 & 0.7 & 0.83 & 12.6 & $^b$ &\\  
\object{EROS2-CG-RCB-3} & J175829-3051.3 & none & \_ & \_ & 14.0 & $^c$ &\\ 
\object{V1783 Sgr} & J180450-3243.2 & 1 & 3.3 & 2.61 & 10.7  & 16-07-2010 & 31-08-1994\\
\object{WX CrA}  & J180850-3719.7 & 2 & 3.5 & 1.94 & 10.5 & 17-07-2010 & 17-08-1993\\
\object{V739 Sgr} & J181311-3016.2 & none & \_ & \_ & 12.9 & 16-07-2010 & 13-06-1990\\
\object{V3795 Sgr} & J181325-2546.9 & 1 & 1.2 & 1.20 & 11.0 & 16-07-2010 & 10-06-1994\\
\object{VZ Sgr}  & J181509-2942.5  & 2 & 4.3 & 1.12 & 10.3 & 17-07-2010 & 10-06-1994\\
\object{RS Tel}   & J181851-4632.9  & 4 & 3.4 & 0.22 & 9.85 & 16-07-2010 & 10-06-1994\\
\object{GU Sgr}  & J182416-2415.4 &  2 & 4.2 & \_ & 10.4 & 23-07-2011 & 11-06-1994\\
\object{V348 Sgr} & J184020-2254.5 & 6 & 3.0 & \_ & 12.2 & 15-07-2010 &\\
\object{MV Sgr}  & J184432-2057.2  &  none  & \_ & \_ & 13.5 & 15-07-2010 &\\
\object{FH Sct}   & J184515-0925.5  &  3 & 2.0 & 1.02 & 12.5 & 24-07-2011 & 29-06-1989\\
\object{V CrA}    & J184732-3809.6  & 1 & 2.8 & 0.54 & 9.7 & 23-07-2011 & 11-06-1994\\
\object{SV Sge}   & J190812+1737.7  & 2 & 2.3 & 2.62 & 10.5 & 24-07-2011 &\\
\object{V1157 Sgr}  & J191012-2029.7  & 9 & 2.8 & 2.17 & 11.5 & 15-07-2010 & 30-08-1994\\ 
 & & & & & & 24-07-2012 &\\
\object{RY Sgr}  & J191633-3331.3 & 2 & 6.0  & \_ & 6.3  & 15-07-2010 & 17-06-1990\\
\object{ES Aql}  & J193222-0011.5  &  10 & 3.2 & 2.05 &  11.65 & 24-07-2011 &\\
\object{U Aqr}   & J220320-1637.6 & 2 & 3.8 & 2.41 & 11.2 & 15-07-2010 & 16-06-1990\\
\hline
\multicolumn{8}{l}{$^a$ \object{RZ Nor} was in a decline phase during our observation. It is worth mentioning that there exists a nearby}\\
\multicolumn{8}{l}{\hspace{15 mm}A star (V$\sim$15 mag). The presence of this blue star explains the unexpected change of colour in RZ Nor during a decline.}\\
\multicolumn{8}{l}{$^b$ See \citet{2011A&A...529A.118T}. $^c$ See \citet{2008A&A...481..673T}. }\\
\multicolumn{8}{l}{Note 1: \object{DY Per}, \object{XX Cam}, \object{Z UMI}, \object{NSV 11154}, \object{V482 Cyg} and \object{UV Cas} are outside the sky area monitored by the ASAS-3 south survey.} \\
\multicolumn{8}{l}{Note 2: The 17 following known Galactic RCBs are too faint to have been catalogued and monitored by the ASAS-3 south survey: } \\
\multicolumn{8}{l}{\hspace{15 mm}MACHO-135.27132.51, -301.45783.9, -308.38099.66 and -401.48170.2237 from \citet{2005AJ....130.2293Z},} \\
\multicolumn{8}{l}{\hspace{15 mm}EROS2-CG-RCB-1 and -4 to -14 from \citet{2008A&A...481..673T} and OGLE-GC-RCB-1 from \citet{2011A&A...529A.118T}.} \\
\hline
\end{tabular}
\end{table*}

\begin{table*}
\caption{Newly discovered and confirmed RCB and DY Per stars
\label{tab.NewRCB}}
\medskip
\centering
\begin{tabular}{lcclcl}
\hline
\hline
Name & ASAS-3 & Coordinates & Other  & Spectrum & Found with \\ 
& id & ($\mathrm{J_{2000}}$) & names & date & analysis?$^a$\\
\hline
\object{ASAS-RCB-1} & J154425-5045.0 & 15:44:25.08 -50:45:01.2  & \object{V409 Nor}, \object{CGCS 3628} & 15-07-2010 & A,B\\
\object{ASAS-RCB-2} & J164124-5147.8 & 16:41:24.74 -51:47:43.4  & \object{IRAS 16375-5141} & 15-07-2010 & A,C,D\\
\object{ASAS-RCB-3}$^d$ & J165444-4926.0 & 16:54:43.60 -49:25:55.0  &	\object{CGCS 3744}, \object{IRAS 16509-4921} & 15-07-2010 & A,B,C,D\\
& & & & 17-07-2010 & \\
\object{ASAS-RCB-4}$^d$ & J170542-2650.0 & 17:05:41.25 -26:50:03.4  & \object{GV Oph} & 15-07-2010 & A,B,C,D\\
\object{ASAS-RCB-5} & J175226-3411.4 & 17:52:25.50 -34:11:28.2  & \object{IRAS 17491-3410} & 15-07-2010 & A,B,C,D \\
& & & & 17-07-2010 & \\
\object{ASAS-RCB-6}$^d$ & J203005-6208.0 & 20:30:04.96 -62:07:59.2  & \object{AN 141.1932} & 28-11-2009 & A,B,C,D\\ 
& & & & 12-12-2010 & \\
& & & & 25-07-2011 & \\
\object{ASAS-RCB-7}$^c$ & J174916-3913.3 & 17:49:15.66 -39:13:16.6  & \object{V653 Sco} & 15-07-2010 & A,D\\
\object{ASAS-RCB-8} & J190640-1623.9 & 19:06:39.87 -16:23:59.2  &  & 15-07-2010 & A\\ 
\object{ASAS-RCB-9}$^{c,d}$ & J162229-4835.9 & 16:22:28.83 -48:35:55.8   & \object{IO Nor} & 15-07-2010 & B,C,D\\ 
\object{ASAS-RCB-10} & J171710-2043.3 & 17:17:10.22 -20:43:15.8  & \object{TYC 6245-382-1} & 15-07-2010 & B,C,D\\ 
\object{ASAS-RCB-11} & J181204-2808.6 & 18:12:03.70 -28:08:36.2  & \object{CGCS 3965}, \object{IRAS 18089-2809} & 15-07-2010 & B,C,D\\ 
\object{ASAS-RCB-12} & J170102-5015.6 & 17:01:01.41 -50:15:34.9  & \object{NSV 8092} & 20-07-2011 & B,C,D\\ 
\object{ASAS-RCB-13}$^c$ & J182443-4524.7 & 18:24:43.46 -45:24:43.8  & \object{V581 CrA} & 10-05-2012 & D\\
\object{ASAS-RCB-14} & J164729-1525.3 & 16:47:29.74 -15:25:22.92 & \object{IRAS 16446-1520} & 06-06-2012 & C,D \\
\object{ASAS-RCB-15} & J170827-3226.8 & 17:08:27.19 -32:26:49.56 & \object{Terz V 3526}  & 06-06-2012 & C,D \\
\object{ASAS-RCB-16} & J171414-2126.1 & 17:14:14.45 -21:26:13.74 & \object{CGCS 379}, \object{IRAS 17112-21220} & 06-06-2012 & C,D \\
\object{ASAS-RCB-17} & J171744-2937.9 & 17:17:44.49 -29:38:00.14 & \object{Terz V 21} & 06-06-2012 & C,D \\
\object{ASAS-RCB-18} & J190009-0202.9 & 19:00:09.44 -02:02:57.88 & \object{IRAS 18575-0207} & 07-06-2012 &  C,D \\
\object{ASAS-RCB-19} & J190133+1456.1 & 19:01:33.68 +14:56:09.61 & \object{NSV 11664}, \object{IRAS 18592+1451}  & 07-06-2012 & D \\
 &  &  &   & 01-08-2012 & D \\
\object{ASAS-RCB-20} & J195343+1441.1 & 19:53:43.15 +14:41:09.34 & \object{PT Aql}, \object{IRAS 19513+1433}  & 01-08-2012 & D \\
\object{ASAS-RCB-21} & J185841-0220.2 & 18:58:41.80 -02:20:11.30 & \object{CGCS 6764}, \object{IRAS 18560-0224} & 23-07-2012 & D \\
\hline
\multicolumn{6}{c}{Other RCB stars spectroscopically confirmed}\\ 
\hline
\object{IRAS 1813.5-2419}$^b$ & 181639-2418.7 & 18:16:39.20 -24:18:33.4 & OGLE-II DIA BUL-SC13 17058 & 17-07-2010 &  \\
\object{V391 Sct} & 182807-1554.9 & 18:28:06.61 -15:54:44.1 & & 15-07-2010 &  \\ 
& & & & 10-05-2012 & \\
\hline
\multicolumn{6}{c}{DY Per stars}\\ 
\hline
\object{ASAS-DYPer-1} & 091802-5402.5 & 09:18:01.59 -54:02:27.4 & \object{V487 Vel} & 17-07-2010 & \\ 
& & & & 11-12-2010 & \\ 
\object{ASAS-DYPer-2} & 122658-6824.3 & 12:26:57.96 -68:24:15.0 & \object{CGCS 3238} & 28-11-2009 & \\ 
& & & & 12-12-2010 & \\
\hline
\multicolumn{6}{l}{$^a$ These new RCBs were discovered using analysis A, B, C and/or D. See the text for details on each. }\\ 
\multicolumn{6}{l}{$^b$ URL: http://ogledb.astrouw.edu.pl/$\sim$ogle/photdb/getobj.php?field=BUL\_SC13\&starid=17058\&db=DIA\&points=good } \\
\multicolumn{6}{l}{$^c$ Also reported as RCB stars by \citet{2012A&A...539A..51T} ; $^d$ Also reported as RCB stars by \citet{2012ApJ...755...98M}} \\
\end{tabular}
\end{table*}

\begin{table*}
\caption{DENIS and 2MASS magnitudes 
\label{tab.IR}}
\medskip
\centering
\begin{tabular}{lclclclcclclclc}
\hline
\hline
 & \multicolumn{7}{c}{ DENIS} & \multicolumn{7}{c}{2MASS} \\ 
Name & I & $\mathrm{\sigma_I}$ & J  & $\mathrm{\sigma_J}$ & K & $\mathrm{\sigma_K}$  & Epoch$^a$ & J &$\mathrm{\sigma_J}$ & H & $\mathrm{\sigma_H}$ & K & $\mathrm{\sigma_K}$ & Epoch  \\ 
 & & & & & & & MJD & & & & & & & MJD \\
\hline
ASAS-RCB-1 &   12.82 &  0.06 &   9.91 &  0.04 &   7.18 &  0.07 & 51276.7998 & 8.83 & 0.03 & 7.70 & 0.05 & 6.58 & 0.01 & 51341.6356\\ 
   	   &     13.19 &  0.02 &  10.23 &  0.04 &   7.39 &  0.07  & 51268.8477 & & & & & & & \\
ASAS-RCB-2 &     10.08 &  0.03  &  8.24  & 0.09  &  6.38  & 0.11 & 52008.7504 & 12.45 & 0.02 & 9.74 & 0.02 & 7.59 & 0.01 & 51347.6877 \\
   	   &	 12.05  & 0.02  &  9.58 &  0.06 &   6.92 &  0.08  & 51655.8479 & & & & & & &   \\ 
ASAS-RCB-3 & & & & & & & & 8.43 & 0.01 & 7.91 & 0.02 & 7.48 & 0.02 & 51701.5563  \\
ASAS-RCB-4 &     11.09 &  0.10 &  10.18 &  0.11 &   9.13  & 0.12 & 51730.6392 & 10.01 & 0.02 & 9.28 & 0.02 & 8.46 & 0.02 & 51004.6576 \\
	   &      9.94 &  0.03 &    8.91  & 0.05 &   7.96 &  0.08  & 51749.5922 & & & & & & &   \\
ASAS-RCB-5 &     11.58 &  0.06 &   9.55 &  0.11  &  7.67 &  0.08  & 51405.5699 & 8.38 & 0.02 & 7.59 & 0.05 & 6.82 & 0.01 & 51040.4850 \\  
ASAS-RCB-6 &     15.82 &  0.06 &  14.37 &  0.13 &  11.89 &  0.11 & 51003.7462 & 11.73 & 0.02 & 11.20 & 0.02 & 10.40 & 0.02 & 51701.8785\\ 
ASAS-RCB-7 & & & & & & & & 12.15 & 0.02 & 10.81 & 0.02 & 9.35 & 0.02 & 51764.4863\\
ASAS-RCB-8 &     10.20 &  0.04 &   9.89 &  0.06  &  9.55  & 0.07 & 51745.7379  & 10.00 & 0.02 & 9.85 & 0.02 & 9.72 & 0.02 & 51761.5412\\
ASAS-RCB-9 &      9.10 &  0.04 &   6.90 &  0.09  &  5.40  & 0.13 & 51387.5397 &  7.26 & 0.02 & 6.47 & 0.02 & 5.74 & 0.01 & 51347.5414\\
	   &	  8.98 &  0.05 &   7.13 &  0.08  &  5.45  & 0.15 & 51395.4921 &  & & & & & &  \\ 
ASAS-RCB-10 &    10.70 &  0.07 &   9.92 &  0.09 &   9.44  & 0.09 & 52048.7917 & 9.62 & 0.02 & 9.28 & 0.02 & 8.80 & 0.02 & 51305.7354 \\
ASAS-RCB-11 & & & & & & & & 9.09 & 0.01 & 8.42 & 0.02 & 7.77 & 0.02 & 51816.5014\\
ASAS-RCB-12 &    14.50 &  0.05 &  12.26  & 0.08  &  9.14  & 0.07 & 51645.8802 & 13.97 & 0.03 & 11.41 & 0.03 & 9.35 & 0.02 & 51355.5598  \\
	    &	 14.01 &  0.04 &  11.91 &  0.07 &   9.05  & 0.06 & 51653.8495 & & & & & & &   \\
ASAS-RCB-13 & & & & & & & & 7.61 & 0.01 & 6.62 & 0.03 & 5.54 & 0.01 & 51316.8659\\
ASAS-RCB-14 &    11.90 &  0.02 &  10.17  & 0.07  &  7.88  & 0.06 & 51394.5565 & 10.20 & 0.02 & 9.06 & 0.02 & 7.85 & 0.01 & 51294.7585 \\
ASAS-RCB-15 & & & & & & & & 14.29 & 0.03 & 11.64 & 0.02 & 9.56 & 0.02 & 51005.6696\\
ASAS-RCB-16  &   10.68 &  0.04 &            &          &          &          & 52071.6501 & 9.78 & 0.02 & 9.04 & 0.02 & 8.48 & 0.02 & 51415.5321  \\
	          &	10.81 &  0.04 &   9.39  & 0.07  &  8.26 &  0.25 & 51754.5688 &         &         &        &          &         &         &                      \\
ASAS-RCB-17 &    12.94  & 0.02 &  10.86  & 0.05 &   9.15 &  0.06 & 51334.7639 & 9.55 & 0.02 & 8.93 & 0.02 & 8.39 & 0.02 & 51006.5688 \\
ASAS-RCB-18 &    10.96 &  0.02 &   8.99  & 0.07  &  7.39 &  0.07 & 51751.6073 & 8.90 & 0.02 & 8.17 & 0.04 & 7.36 & 0.02 & 51706.6558 \\ 
ASAS-RCB-19 & & & & & & & & 15.30 & 0.05 & 12.41 & 0.02 & 9.99 & 0.02 & 50641.8467\\
ASAS-RCB-20 & & & & & & & & 15.31 & 0.05 & 12.41 & 0.03 & $>$9.95 & \_ & 50941.9517\\
ASAS-RCB-21 & & & & & & & & 8.91 & 0.02 & 7.75 & 0.03 & 6.51 & 0.02 & 51703.6733\\
IRAS1813.5-2419 & 11.53 & 0.03 & 9.73 &  0.06  &  8.02 &  0.07 & 51342.7916 & 10.82 & 0.05 & 9.75 & 0.05 & 8.63 & 0.03 & 51816.5180  \\
V391 Sct	& 11.83 & 0.02 & 10.53 & 0.07 & 9.77 & 0.13 & 51001.6767 & 10.43 & 0.03 & 9.80 & 0.04 & 9.04 & 0.03 & 51365.6109  \\ 
\hline
\multicolumn{15}{c}{ DY Per  stars} \\
\hline
ASAS-DYPer-1 & 9.25 & 0.03 & 5.55 & 0.07 & 3.79 & 0.08 & 51544.6687 & 5.75 & 0.02 & 4.12 & 0.24 & 3.18 & 0.28 & 51550.8476\\
ASAS-DYPer-2 & 9.43 & 0.04 & 7.63 & 0.06 & 5.82 & 0.09 & 51236.8090 & 7.64 & 0.02 & 6.52 & 0.03 & 5.98 & 0.02 & 51922.8383\\
\hline
\multicolumn{15}{l}{$^a$ We note that the epochs listed in the DENIS catalogue are not reliable: we used the epoch listed in the headers of the original images.}\\
\hline
\end{tabular}
\end{table*}

\begin{table*}
\caption{Variability of the newly discovered and confirmed RCB stars
\label{tab.NewRCBvar}}
\medskip
\centering
\begin{tabular}{lr@{.}lr@{.}lcccc}
\hline
\hline
Name & \multicolumn{4}{c}{Galactic coordinates} & Number of declines & $\mathrm{V_{ASAS,max}}$ & $\mathrm{E(B-V)}^a$ & $\mathrm{A_V}$ (using $\mathrm{R_V}\sim3.1$) \\ 
    & \multicolumn{2}{c}{l (deg)} & \multicolumn{2}{c}{b  (deg)} & observed & (mag) & & (mag)\\
\hline
ASAS-RCB-1 & 328&66525 & 3&27132 & 1 & 11.9 & 1.36 & 4.22 \\
ASAS-RCB-2 & 334&43866 & -3&58426 & 3 & 11.8$^{\star}$ & 0.80 & 2.48 \\
ASAS-RCB-3 & 337&61260 & -3&68098 & 2 & 11.8 & 0.89 & 2.76 \\
ASAS-RCB-4 & 356&81951 & 8&49585 & 3 & 11.9 & 0.32 & 0.99 \\
ASAS-RCB-5 & 356&23379 & -3&94349 & 5 & 12.3 & 0.81 & 2.51 \\
ASAS-RCB-6 & 334&26108 & -35&21748 & 1 & 13.0 & 0.05 & 0.16 \\
ASAS-RCB-7 & 351&55769 & -5&94405 & 4 & 12.5$^{\star}$ & 0.42 & 1.30 \\
ASAS-RCB-8 & 19&86093 & -10&68582 & 1 & 10.9 & 0.20 & 0.62 \\
ASAS-RCB-9 & 334&70421 & 0&81022 & 5 & 10.8 & 5.56 & $>5^b$ \\
ASAS-RCB-10 & 3&41300 & 9&87397 & 1 & 11.4 & 0.61 & 1.89 \\
ASAS-RCB-11 & 3&58229 & -4&64083 & 4 & 11.8 & 0.52 & 1.61 \\
ASAS-RCB-12 & 337&59120 & -4&99059 & 1 & 11.7$^{\star}$ & 0.57 & 1.77 \\
ASAS-RCB-13 & 349&02823 & -14&60072 & 6 & 9.9$^{\star}$ & 0.07 & 0.22 \\
ASAS-RCB-14 & 3&67329 & 18&68830 & 2 & 12.5$^{\star}$ & 0.55 & 1.71 \\
ASAS-RCB-15 & 352&63077 & 4&68055 & 2 & 14.2$^{\star}$ & 0.58 & 1.80 \\
ASAS-RCB-16 & 2&42136 & 10&03462 & 4 & 12.8 & 0.76 & 2.36 \\
ASAS-RCB-17 & 356&08809 & 4&70730 & 3 & 13.1 & 0.92 & 2.85 \\
ASAS-RCB-18 & 32&10038 & -2&87384 & 3 & 13.6 & 1.05 & 3.26 \\
ASAS-RCB-19 & 47&38762 & 4&56290 & 2 & 13.5$^{\star}$ & 0.77 & 2.39 \\
ASAS-RCB-20 & 53&19579 & -6&63163 & 1 & 12.7$^{\star}$ & 0.25 & 0.78 \\
ASAS-RCB-21 & 31&67823 & -2&67951 & 3 & 12.8$^{\star}$ & 1.21 & 3.75 \\
IRAS 1813.5-2419 & 7&45702 & -3&73148 & 0 & 12.6 & 1.48 & 4.59 \\
V391Sct & 16&14758 & -2&18048 & 0 & 13.25 & 1.95 & 6.05 \\
\hline
ASAS-DYPer-1 & 274&77401 & -3&27644 & 1 & 9.4 & 0.89 & 2.76 \\
ASAS-DYPer-2 & 300&67291 & -5&63982 & 1 & 11.95 & 0.33 & 1.02 \\
\hline
\multicolumn{9}{l}{$^a$ From \citet{1998ApJ...500..525S} with 4 nearest pixels interpolation. }\\
\multicolumn{9}{l}{$^b$ The \citet{1998ApJ...500..525S} reddening estimate is overestimated in high-extinction areas \citep[see][Fig.6]{2008A&A...481..673T}.}\\
\multicolumn{9}{l}{$^{\star}$ Not confident that real maximum magnitude reached during ASAS-3 observations.}\\
\hline
\end{tabular}
\end{table*}

\begin{table*}
\caption{Nature of the other ASAS objects spectroscopically followed up and rejected as RCB or DY Per candidates.
\label{tab.Otherstars}}
\medskip
\centering
\begin{tabular}{lcll}
\hline
ASAS-3 id & Coordinates ($\mathrm{J_{2000}}$)  (from ASAS) & SIMBAD name & Nature of the object\\ 
\hline
\object{014428+0536.6} & 01:44:28 +05:36:36 & \object{TYC 35-752-1} & K4III	 \\ 
\object{020504+1925.4} & 02:05:04 +19:25:24 & \object{SW Ari} & Mira, M6	 \\  
\object{020653+1517.7} & 02:06:53 +15:17:42 & \object{TT Ari} & A star + H, HeI, CaII emission	 \\ 
\object{030615-3942.5}$^c$ & 03:06:15 -39:42:30 &  & Mira, M6	 \\  
\object{035412+1810.3} & 03:54:12 +18:10:18 & \object{IRAS 03513+1801} & Mira, M8	 \\  
\object{040907-0914.2} & 04:09:07 -09:14:12 & \object{EV Eri} & 	 C* (Strong 13C at 4744 angs)	 \\  
\object{042841+2150.6} & 04:28:41 +21:50:36 & \object{IRAS 04257+2144}  & Mira, M6	 \\  
\object{043547+2250.4} & 04:35:47 +22:50:24 &  \object{HQ Tau} &	 A star reddened + 	$\mathrm{H\alpha}$ em. \\  
\object{045331+2246.5} & 04:53:31 +22:46:30 & \object{CGCS 808} & 	 C* (Strong 13C at 4744 angs)	 \\ 
\object{050430-0347.2} & 05:04:30 -03:47:12 &  \object{UX Ori} &  A7V + $\mathrm{H\alpha}$ em. 	 \\  
\object{050500+0856.1} & 05:05:00 +08:56:06 & \_ & 	 C* ($\mathrm{H\alpha}$ emission)	 \\  
\object{051314-0844.9} & 05:13:14 -08:44:54 &   &	F8III	 \\   
\object{052114+0721.3} & 05:21:14 +07:21:18 & \object{V1368 Ori} & 	 C* (Strong 13C at 4744 angs)	 \\  
\object{053302+1808.0} & 05:33:02 +18:08:00 &  \object{IRAS 05301+1805} &	 C* (Strong 13C at 4744 angs)	 \\ 
\object{053625-6941.5} & 05:36:25 -69:41:30 & \object{TYC 9167-592-1} & K2III + H, OI, CaII emission\\  
\object{053659-0609.3} & 05:36:59 -06:09:18 & \object{HD 37258}  &	A3V + $\mathrm{H\alpha}$ em.	 \\ 
\object{053713-0635.0} & 05:37:13 -06:35:00 &  \object{BF Ori}  &	 A7III + $\mathrm{H\alpha}$ em.		 \\  
\object{053931+2622.5} & 05:39:30 +26:22:30 &  \object{RR Tau}  &	  A star reddened + 	$\mathrm{H\alpha}$, CaII em.	 \\ 
\object{054012-0942.2} & 05:40:12 -09:42:12 & \object{V350 Ori}  &  A7V + $\mathrm{H\alpha}$ em. 	 \\ 
\object{054339-0504.1} & 05:43:39 -05:04:00 & \object{DM Ori} &  G star reddened + $\mathrm{H\alpha}$ em. 	 \\ 
\object{054635+2538.1} & 05:46:35 +25:38:06 & \object{CGCS 1049} & C* (Strong 13C at 4744 angs)	 \\  
\object{055343-1023.9} & 05:53:43 -10:24:00 & \object{V1818 Ori}  &	 A star reddened + $\mathrm{H\alpha}$, CaII em. \\   
\object{062348-2051.5} & 06:23:48 -20:51:30 & \object{TYC 5946-114-1} & G0III + $\mathrm{H\alpha}$ em.	 \\ 
\object{064107+1026.7} & 06:41:07 +10:26:42 & \object{SS Mon} &	K7V + $\mathrm{H\alpha}$ em., PMS? \\ 
\object{064854-4249.4} & 06:48:54 -42:49:24 & \object{IRAS 06473-4246}  & Mira M9  \\ 
\object{070615-1904.4} & 07:06:15 -19:04:24 & \_  & A0V + $\mathrm{H\alpha}$ em.  \\  
\object{070813-1036.0}$^a$ & 07:08:13 -10:36:00 &  \object{ALS 181} &	 A star reddened 	 \\  
\object{071204+2706.3} & 07:12:04 +27:06:18 & \object{TYC 1904-1095-1} & Mira, M6	 \\  
\object{071839-0117.0} & 07:18:39 -01:17:00 &  & C* ($\mathrm{H\alpha}$ emission)	 \\  
\object{072221-5049.4} & 07:22:21 -50:49:24 & \object{NY Pup}  & Mira, M6	 \\ 
\object{072419+0912.6} & 07:24:19 +09:12:30 & \object{GSC 00764-00509} &   Giant,  G? reddened	 \\  
\object{075752-0916.0} & 07:57:52 -09:16:00 & \object{IRAS 07554-0907} &	 C* ($\mathrm{H\alpha}$, $\mathrm{H\beta}$, $\mathrm{H\gamma}$ + $\mathrm{H\delta}$ em.)	 \\ 
\object{081204-2002.4} & 08:12:04 -20:02:24 & \object{V535 Pup}  &	 C* (Strong 13C at 4744 angs)	 \\  
\object{081447-4441.6} & 08:14:47 -44:41:36 & \object{TYC 7672-563-1} &	G0III + $\mathrm{H\alpha}$ em. \\  
\object{083236-3759.0} & 08:32:36 -37:59:00 & \object{FX Vel} &	A3III + $\mathrm{H\alpha}$, CaII, OI, OIII em.	 \\  
\object{095221-4329.7}$^c$ & 09:52:21 -43:29:42 & \object{TYC 7706-560-1} & Mira, M6	 \\ 
\object{100731-0918.1} & 10:07:31 -09:18:06 & \object{BD-08 2852} &  M3III	 \\  
\object{110204-6209.7} & 11:02:04 -62:09:42 & \object{V802 Car} & F2III\\  
\object{110815-7733.9} & 11:08:15 -77:33:54 & \object{HP Cha}  & K7V + $\mathrm{H\alpha}$ em., PMS?	 \\ 
\object{111846-5612.6}$^a$ & 11:18:46 -56:12:36 &  & A7V \\  
\object{120729+0036.9} & 12:07:29 +00:36:54 &  & K2III + $\mathrm{H\alpha}$ P-cygne em. \\  
\object{122112-4912.7} & 12:21:12 -49:12:42 & \object{SX Cen} &  B5III	 \\ 
\object{133504-6345.9} & 13:35:04 -63:45:54 & \object{TYC 8999-807-1} & F0III + $\mathrm{H\alpha}$ em.\\  
\object{141600-6153.9} & 14:16:00 -61:53:54 & \object{V417 Cen}  &	F6V reddened + $\mathrm{H\alpha}$ em.	 \\ 
\object{144959-6920.9} & 14:49:59 -69:20:54 & \object{BL Cir} & Mira, M6\\
\object{150323-6323.0} & 15:03:23 -63:23:00 & \object{DG Cir}  & A star + nebulae	 \\  
\object{152008-6148.5} & 15:20:08 -61:48:24 &  &  F6V	+ Pcygne $\mathrm{H\alpha}$\\  
\object{152604-7004.0} & 15:26:04 -70:04:00 & \object{IRAS 15211-6953}  &  Mira, M7 \\ 
\object{153213-2854.3} & 15:32:13 -28:54:18 & \object{BX Lib} &	K7.5III + $\mathrm{H\alpha}$ em. \\  
\object{155517-2924.6}$^c$ & 15:55:17 -29:24:36 &  \object{V1317 Sco} &	Mira, M6	 \\  
\object{164557-4330.1} & 16:45:57 -43:30:06 &  \_ & G0III + $\mathrm{H\alpha}$ P-cygne em.  \\  
\object{165449-4032.8} & 16:54:49 -40:32:48 &  \_ & F8V \\  
\object{165452-3037.3}$^c$ & 16:54:52 -30:37:18 & \object{CL Sco} &	 Symbiotic star with M giant, M4	 \\  
\object{165750-3234.7}$^c$ & 16:57:50 -32:34:42 & \object{EW Sco} &  Mira, M6	 \\ 
\_ & 17:17:22 -20:22:38 & \object{V1773 Oph} & F2III, binary?\\
\object{172109-4644.1}$^c$ & 17:21:09 -46:44:06 & \object{IRAS 17174-4641} & Mira, M7	 \\  
\object{173540+1535.2} & 17:35:40 +15:35:12 & \object{MW Her} &	Mira, M10	 \\ 
\object{173708-4708.1} & 17:37:08 -47:08:00 &  & M0III	 \\  
\object{173729-3743.4} & 17:37:29 -37:43:24 &  &	F8IV reddened \\  
\object{174600-2321.2}$^c$ & 17:46:00 -23:21:12 &  & F0I reddened \\ 
\object{175013-0642.5}$^c$ & 17:50:13 -06:42:30 & \object{Oph 1898} &  K7III + H, HeI, CaII, OI em. (Nova) \\  
\object{175335-3805.0} & 17:53:34 -38:05:00 & \object{V383 Sco} &	 Blend F8I +  M6(I to III) +  $\mathrm{H\alpha}$ em.\\  
\object{180204-2337.7} & 18:02:04 -23:37:42 & \object{V5097 Sgr} &	Wolf Rayet, WC9 + 2 mag. declines\\ 
\_ & 18:08:36 -15:04:00 & \object{GM Ser} & Mira, M7 \\
\object{181448-1840.3}$^{a,b}$ & 18:14:48 -18:40:18 & \object{2MASS J18144821-1840219}  & F0I reddened +  $\mathrm{H\alpha}$ em.\\ 
\hline
\multicolumn{4}{l}{$^a$ The ASAS light curves of these objects show some variability that are certainly not real but due to some photometric systematics.}\\
\multicolumn{4}{l}{Nevertheless, we followed-up these objects for completeness in our study.}\\
\multicolumn{4}{l}{$^b$ The star that was followed-up spectroscopically is 2MASS18144821-1840219. This is the star selected by our IR colour criteria.}\\ 
\multicolumn{4}{l}{$^c$ These ASAS objects were selected from the criteria defined to search for DY Per stars (see Sect. \ref{ana_dyper}) }.\\ 
\hline
\end{tabular}
\end{table*}

\begin{table*}
\caption{Nature of the other ASAS objects spectroscopically followed up and rejected as RCB or DY Per candidates (continued).
\label{tab.Otherstars2}}
\medskip
\centering
\begin{tabular}{lcll}
\hline
ASAS-3 id & Coordinates ($\mathrm{J_{2000}}$)  (from ASAS) & SIMBAD name & Nature of the object\\ 
\hline
\object{181706-2358.0} & 18:17:06 -23:58:00 & \object{V2331 Sgr} & SC star (weak TiO and weak CN)\\
\object{182128-2445.1} & 18:21:28 -24:45:06 & \object{V1860 Sgr} & G6III reddened \\
\object{182253-2336.9} & 18:22:53 -23:36:54 & \object{NSV 10706} &	 G5IV	 \\ 
\object{182502-2308.0} & 18:25:02 -23:08:00 & \object{V3816 Sgr} & G5III	 \\ 
\object{182715-2511.9} & 18:27:15 -25:11:54 & \_  & Mira, M7 \\ 
\object{182726-0434.8} & 18:27:26 -04:34:48 & \object{2MASS J18272608-0434473} &	 G5III reddened + $\mathrm{H\alpha}$ em.\\  
\object{182848+0008.7} & 18:28:48 +00:08:42 & \object{VV Ser} &	A5V reddened +  $\mathrm{H\alpha}$ em.	 \\ 
\object{190740-1535.0} & 19:07:40 -15:35:00 &  \object{TYC 6283-1454-1}  & Mira, M6	 \\  
\object{190842-4322.1} & 19:08:42 -43:22:06 & \_  & Mira, M8	 \\  
\object{192613+0501.5} & 19:26:13 +05:01:30 & \_ &	M0I + $\mathrm{H\alpha}$ em., No MgI,NaI,CaII lines	 \\  
\object{194714+1929.3} & 19:47:14 +19:29:18 & \object{RZ Vul} &	 G5III reddened\\  
\object{195538-0018.7} & 19:55:38 -00:18:42 & \object{V560 Aql}  & Mira, M9	 \\ 
\object{200055+2305.7} & 20:00:55 +23:05:42 & \object{CF Vul} &  G?III  + weak carbon features	 \\  
\object{201519+2603.6} & 20:15:19 +26:03:36 &  \object{EF Vul} &	G8V + $\mathrm{H\alpha}$ em.	 \\  
\hline
\end{tabular}
\end{table*}

\section{Individual stars \label{sec_stars}}

In this section, we discuss each newly discovered RCB and DY Per star. Their DENIS and 2MASS IR magnitudes are listed in Table~\ref{tab.IR} and their recorded ASAS-3 maximum V magnitudes are given in Table~\ref{tab.NewRCBvar} with an estimation of their reddening. Unfortunately, the DENIS and 2MASS surveys were completed at the time when the ASAS-3 south survey started and therefore we do not know the brightness phase of the 23 new RCB stars at those epochs. Interestingly, we note that 6 out of the 23 new RCB stars were already catalogued as carbon stars in the literature. They are ASAS-RCB-1, -3, -11, -16, -20 and -21.

\begin{itemize}
\item \textbf{ASAS-RCB-1}: Also known as V409 Nor, it was classified for the first time as a carbon star in the cool carbon stars catalogue of \citet{1973PW&SO....1...1S}.  It was recently recognized as a probable RCB star by \citet{2011PZ.....31....4K}. Its spectrum shows strong Ca-II absorption lines indicating that it is a warm RCB star. Only one decline of $\sim$2 mag is observable in its ASAS-3 light curve.
 
\item \textbf{ASAS-RCB-2}: Has a high IR excess, with $\mathrm{J-K}>4$ mag, indicating the presence of a bright hot circumstellar shell. It remained fainter than 15th magnitude for more than 3.5 years, between 2001 and 2004. 
 
\item \textbf{ASAS-RCB-3}: Also known as CGCS 3744, it was classified for the first time as a carbon star in the cool carbon stars catalogue of \citet{1973PW&SO....1...1S}. It has also recently been found to be an RCB star by \citet{2012ApJ...755...98M}. 
 
\item \textbf{ASAS-RCB-4}: Also known as GV Oph, it has recently been found to be an RCB star by \citet{2012ApJ...755...98M}.
 
\item \textbf{ASAS-RCB-5}: The only newly discovered RCB that has been monitored during phase 2 of the ASAS survey. The I band light curve covers a time range between HJD$\sim$2450500 and  2451500, and shows one decline. It reaches a maximum magnitude of $\mathrm{I_{max}}\sim 10$ mag. 
 
\item \textbf{ASAS-RCB-6}: Identified as a variable star by \citet{1932AN....246..437L}. \citet{2012ApJ...755...98M} has also found recently that it is indeed an RCB star. Interestingly, its spectrum shows features due to C$_2$ but none due to CN, and weak Ca-II absorption lines indicating a low effective temperature.

\item \textbf{ASAS-RCB-7}: Also known as V653 Sco, it was misclassified as a Mira star in the General Catalog of Variable Stars (GCVS). It was reported as a variable star for the first time by \citet{1971GCVS3.C......0K} with a total amplitude larger than 2 mag.
 
\item \textbf{ASAS-RCB-8}: Only one decline was observed during the 10-year long ASAS-3 survey. With 0.003 mag.day$^{-1}$, the decline rate is about 10 times slower than to classical RCB stars.  Its spectrum is similar to the warm RCB star UX Ant.  
 
\item \textbf{ASAS-RCB-9}: Also known as IO Nor, it was recently identified as an RCB star by \citet{2012ApJ...755...98M}. It was misclassified as a Mira star in the GCVS. We note that ASAS-RCB-9 was identified as a Tycho-2 star having an infrared excess in the MSX Point Source Catalogue \citep{2005MNRAS.363.1111C}. With $\mathrm{V_{max}}\sim 10.8$ mag, it is the brightest of all the new RCB stars reported in this article.
  
\item \textbf{ASAS-RCB-10}: Its ASAS-3 light curve shows only one large and rapid decline of $\sim$2.5 mag around MJD$\sim$53700 days. Its spectrum is similar to the warm RCB star UX Ant, but affected by strong dust reddening.  
 
 \item \textbf{ASAS-RCB-11}: Also known as CGCS 3965, it was classified for the first time as a carbon star in the cool carbon stars catalogue of \citet{1973PW&SO....1...1S}. Its ASAS-3 light curve shows four decline phases in 10 years and some photometric oscillation at maximum brightness. 

 \item \textbf{ASAS-RCB-12}: It was recently recognized as a probable RCB star by \citet{2011PZ.....31....6K}. It has a high IR excess, with $\mathrm{J-K}>4$ mag, indicating the presence of a bright hot circumstellar shell. From its ASAS-3 light curve, ASAS-RCB-12 seems to have been in a decline phase for more than 5 years and slowly recovered to maximum brightness in $\sim$2010. Its spectrum shows strong features due to C$_2$ but only very weak features due to CN. 
 
 \item \textbf{ASAS-RCB-13}: Also known as V581 CrA, it was listed as a variable star by \citet{1968IBVS..311....1K}. It was identified as a probable RCB star by \citet{2012A&A...539A..51T} because it was found to be misclassified as a Mira variable in the GCVS, which did not match its WISE mid-infrared colour. Its spectrum shows weak CN features compared to the C$_2$.

\item \textbf{ASAS-RCB-14}: Has also been detected by the IRAS satellite. Its ASAS-3 light curve has only a small number of measurements around MJD$\sim$53150 days, when ASAS-RCB-14 was recovering its brightness. It reaches a maximum magnitude of  $\mathrm{V_{max}}\sim 13.2$  before undergoing decline for a long period, indicating a phase of high dust production activity.

\item \textbf{ASAS-RCB-15}: Has been identified as a variable star by \citet{1997A&AS..123..507T}. We found that it has a high IR excess, with $\mathrm{J-K}>4$ mag, indicating the presence of a bright hot circumstellar shell. Its ASAS-3 light curve has only a small number of measurements showing a recovery phase followed by a decline around MJD$\sim$54550 days. It reaches a maximum magnitude of $\mathrm{V_{max}}\sim 14.3$ mag. ASAS-RCB-15 demonstrates the success of this study in extending the limits of detectability.
 
\item \textbf{ASAS-RCB-16}: Also known as CGCS 3790, it was classified for the first time as a carbon star in the cool carbon stars catalogue of \citet{1973PW&SO....1...1S}. It was also identified as a variable star by \citet{1968BANS....2..293P}. We can be confident that we measured its real maximum magnitude, $\mathrm{V_{max}}\sim 12.8$ mag, as it lasted about 1000 days.
  
\item \textbf{ASAS-RCB-17}: Was already identified as a variable star by \citet{1982A&AS...49..715T}. One fast decline of $\sim$1.2 mag is observable on its ASAS-3 light curve at MJD$\sim$52700 days. We can be confident that we measured its real maximum magnitude, $\mathrm{V_{max}}\sim 13.1$ mag, as it lasted about 700 days.
   
\item \textbf{ASAS-RCB-18}: Its ASAS-3 light curve presents few variations with an amplitude of $\sim$1 mag, but remains overall relatively stable as it does not undergo large decline phases. We selected this star for spectroscopic follow-up because of the fast decline observed at MJD$\sim$53600 days. It was selected in the first place because of its mid IR WISE colours and 2MASS IR excess indicating the presence of a hot circumstellar shell. It has also been detected by the IRAS survey. The measurement at $\sim$65 $\mu$m suggests that there exists a cold remnant around ASAS-RCB-18 that may be related to an old evolutionary phase. A $\sim$80$\pm$30 K blackbody was fitted to this star (see Figure~\ref{fig_SED}).
  
\item \textbf{ASAS-RCB-19}: Also known as NSV 11664, it was identified as a variable star by \citet{1966AN....289....1H}.  It has a high IR excess, with $\mathrm{J-K}>4$ mag, indicating the presence of a bright hot circumstellar shell.

\item \textbf{ASAS-RCB-20}: Also known as PT Aql, it was already identified as a carbon star by \citet{1993A&AS...99...31G}. It was misclassified as a Mira star in the GCVS. We note that it has a high IR excess, with $\mathrm{J-K}>4.5$ mag, indicating the presence of a bright hot circumstellar shell.  It has also been detected by the IRAS satellite.
 
\item \textbf{ASAS-RCB-21}: Was already identified as a carbon star by \citet{1990PNAOJ...1..207S}. It has strong NaI (D) emission lines indicating that its spectrum was taken during a decline phase. Only one rapid recovery phase is observable on its ASAS-3 light curve at the end of the survey, MJD$\sim$55000 days. Due to the quality of its spectrum and the number of measurements in its ASAS-3 light curve ($\mathrm{D_{max}}\sim 1$ mag), we consider that ASAS-RCB-21 requires further confirmation that it is indeed an RCB star. We note that it is not visible on all the plates from the USNO survey and that it has also been detected by the IRAS satellite.
    
\item \textbf{IRAS 1813.5-2419}: Was proposed as an RCB candidate by \citet{2007PZ.....27....7G}, due to the large declines observed in its OGLE-II light curve.
 
\item \textbf{V391 Sct}:  Although listed as a variable star by \citet{1975IBVS..985....1M}, it has been mis-classified  as a cataclysmic variable. V391 Sct is a warm RCB similar to RY Sgr, according to its spectrum. It was originally classified as a dwarf nova (mag = 17-13).  Its ASAS-3 light curve shows that V391 Sct was photometrically stable between 2001 and 2010, with V$\sim$13 mag. Plate analysis shows that it was very faint in 1952 and 1963 (Brian Skiff, private discussion) .

\item \textbf{ASAS-DYPer-1}: Was first classified as a carbon star by \citet{1971A&AS....4...51W}. Its ASAS-3 light curve shows a recovery phase of $\sim$3.8 mag followed by large amplitude oscillations ($\sim$0.8 mag peak to peak).  We note also that the $^{13}$C$^{12}$C band-head at 4740 \AA{} indicates that the atmosphere of ASAS-DYPer-1 is rich in $^{13}$C.  This star was classified as a possible RCB star in the GCVS. We note also that we observed ASAS-DYPer-1 to have a circumstellar shell with a relatively cold temperature of $\sim$500 K (see Figure~\ref{fig_SED}). That is the coldest shell recorded for a DYPer star. Finally, we note that with K$\sim$3.2 mag,  ASAS-DYPer-1 is at least 1.2 mag brighter that the prototype DY Per star, DY Persei.

\item \textbf{ASAS-DYPer-2}: Was first classified as a carbon star in the cool carbon stars catalogue of \citet{1973PW&SO....1...1S}. A small CH (G band) absorption line is observable, which indicates that some hydrogen is present. Its ASAS-3 light curve presents only a small symmetric decline of $\sim$1.4 mag. 

\end{itemize}

\section{Discussion  \label{sec_discu}}

Four different analyses were applied to find new RCB stars in the entire ASAS-3 south dataset, using the ACVS1.1 variable catalogue and the entire ASAS-3 south source catalogue, and by directly interrogating the ASAS-3 light curve database. Candidates were pre-selected by requiring both IR excesses and large light curve variations. The final check was a visual inspection of the light curve.  Only 4 of the 21 new RCBs were discovered in all four analyses and 3 were found by only one analysis: the warm RCB ASAS-RCB-8 by Analysis A, and ASAS-RCB-13 and -19 by Analysis D.

The two classic analyses (A and B), which used criteria based on light curve variability, were limited by the magnitude cutoff of the ASAS-3 survey ($\mathrm{V_{lim}}\sim 14$ mag). Indeed, the detection efficiency was less for fainter RCBs ($V>12$ mag) as no margin was left to detect large declines.

The most efficient method proved to be Analysis D. In this method, the light curves of candidates pre-selected by \citet{2012A&A...539A..51T} using their 2MASS and WISE IR colours, and the subsequent list using the WISE all-sky data release \citep{2012TissInPrepa}, were downloaded directly through the ASAS-3 database. As a result, 17 of the 19 new RCBs were discovered. This confirms the high success rate that can be obtained by searching for new RCB stars using only near- and mid- IR photometry pre-selection. Two out of these 17 RCBs were not catalogued in both ASAS-3 catalogues used, which highlights one limitation of using optical source catalogues to search for RCBs. Such catalogues are limited by the epochs chosen to make the reference images and the number of measurements in the light curve used to validate an object. However, we note that the light curves of these two RCBs were available on the web interface, meaning that they were indeed catalogued at one stage and consequently monitored. This also suggests that some RCBs may still remain undetected among the ASAS-3 images. Only careful source detection, performed on each individual image, can resolve this potential issue.

Overall, the analyses undertaken were useful in finding new RCBs. Together, they achieve a high detection efficiency over a large range of RCB parameters such as luminosity, temperature, dust production activity and shell brightness. Using the already known or newly spectroscopically confirmed bright RCBs listed in tables~\ref{tab.KnownRCB} and \ref{tab.NewRCBvar}, it is possible to estimate an overall detection efficiency  of $\sim$90\% for RCBs brighter than V$\sim$13. The remaining 10\%  are missing mainly due to the fact that some of the bright RCBs did not show any declines on their respective ASAS-3 light curves (i.e., Y Mus, V739 Sgr and IRAS 1813.5-2419). This demonstrates the need for long baseline photometric monitoring.
 
The analysis recently published by \citet{2012ApJ...755...98M} is a promising method of finding RCB stars in future large scale monitoring surveys. \citet{2012ApJ...755...98M} analysed the light curves of each object listed in the ACVS1.1 catalogue using a machine-learned algorithm to classify objects and applied a hard cut on the probability that an object belongs to the RCB class. After the spectroscopic follow-up of their candidates they found 4 new RCBs, which we also found, namely  ASAS-RCB-3, -4, -6 and -9.  It would be very interesting to re-apply their analysis to the entire ASAS-3 south dataset, and also to the ASAS-north dataset. It is hoped that the newly discovered RCBs reported in this article may help to tighten the constraints of their method. We note that detection efficiency will also be substantially increased if the 2MASS and WISE colours are added as parameters, as they reveal the typical warm RCB's circumstellar shell.

\section{Summary}

We discovered 21 new RCB stars and 2 new DY Per stars in the ASAS-3 south datasets, and spectroscopically confirmed 2 previously known RCB candidates, IRAS 1813.5-2419 and V391 Sct. The total number of confirmed RCB stars is now 76 in our Galaxy and 22 in the Magellanic Clouds. We used the ASAS-3 ACVS1.1 variable stars and the ASAS-3 south source catalogues, but also directly interrogated the ASAS-3 light curve database for candidates that were not listed in either of them.  Indeed, the search for RCB stars in an optical dataset is made difficult by the sudden declines in brightness that may last for some weeks or for some years. RCB stars cannot therefore automatically be listed in the reference source catalogue. Furthermore, we note that 5 of the newly discovered RCB stars (ASAS-RCB-2, -12, -15, -19 and -20) have a high IR excess with $\mathrm{J-K}>4.5$ mag. This indicates that there exists a population of RCB stars with a very dense circumstellar shell and high dust production activity, which is confirmed by their light curves. We also note that ASAS-RCB-14, -15, -19, -20 and -21 have only between 10 and 50 measurements in their respective ASAS light curves. Additional photometric follow-up will be needed for these five RCBs in order to detect more aperiodic fast declines and to definitely confirm their nature.

We have demonstrated that the use of the 2MASS and WISE IR catalogues is of great value in the selection of RCB candidates. The search for RCB stars can indeed be undertaken independently of optical monitoring surveys, without losing much detection efficiency ($\sim$10\%). The more time-consuming optical light curve analysis targeting large fast declines in brightness, has the sole advantage of recovering the warmest RCB stars that would otherwise be missed by the IR selections. 

To confirm new RCB stars, it is at this stage still necessary to obtain a long-term light curve showing a drop in brightness as the photosphere is obscured and at least a spectrum indicating an atmosphere with a high carbon content and a hydrogen deficiency. However, we are working toward criteria that would confirm an RCB star based on an infrared excess that indicates an ongoing dust production, and spectroscopic characteristics such as the $^{12}$C/$^{13}$C isotopic relative abundance and the C$_2$/CN ratio \citep{2003MNRAS.344..325M}. These criteria have not yet been clearly defined. The task is complicated by the RCBs' large range of effective temperature (between 4000 and 8000 K), and their diverse atmospheric abundances \citep{2011MNRAS.414.3599J}. For instance, V854 Cen shows strong Balmer lines and a few RCB stars do have detectable $^{13}$C. The stars V CrA, V854 Cen, VZ Sgr, and UX Ant have measured $^{12}$C/$^{13}$C $<$25 \citep{1989MNRAS.238P...1K,2008MNRAS.384..477R,2012ApJ...747..102H}.

In conclusion, we estimate that the overall detection efficiency of our search for RCB stars among the entire ASAS-3 south dataset is about 90\%  for RCBs brighter than V$\sim$13. It is still too early to estimate the total number of RCB stars that exist in our Galaxy and therefore their formation rate. More searches at fainter magnitudes are needed to increase the sample and to get a better picture of the RCB population. However, we found that the Galactic RCB spatial distribution and apparent magnitudes now clearly indicate that RCBs have a bulge-like distribution and are therefore, as expected, an old population of stars. The high interstellar extinction toward the Galactic bulge prevents the detection of more RCBs. This region is exactly where most remaining undetected Galactic RCBs are to be found.

Numerous of those RCB stars should be detected by the ESA's mission Gaia, to be launched in 2013. For 5 years, Gaia will scan the whole sky with a magnitude limit of V$\sim$20 mag and issue near real-time alerts on RCB-type anomalies, providing both light curves and low-resolution spectroscopy for improved objects classification \citep{2012arXiv1210.5007W}.


\begin{acknowledgements}
Patrick Tisserand would like to thank especially Daniel Bayliss and George Zhou for obtaining the spectra of V1860 Sgr, V581 CrA, V391 Sct, V2331 Sgr and V1773 Oph, Mike Bessell for his invaluable assistance with classifying the rejected stars, and last but not least, Tony Martin-Jones, Joseph Falsone, Julia Jane and Barbara Jean Karrer for their careful readings, comments and supports. The authors would like to thank the ASAS collaboration for producing a source catalogue of their south survey. Parts of this research were conducted by the Australian Research Council Centre of Excellence for All-sky Astrophysics (CAASTRO), through project number CE110001020. DW acknowledges support from the Natural Sciences and Engineering Research Council of Canada (NSERC). DK thanks the University of the Western Cape and the South African National Research Foundation (NRF) for continuing financial support. This research has made use of the SIMBAD, Vizier and Aladin databases, operated at CDS, Strasbourg, France.  This research has also made use of the NASA/IPAC Infrared Science Archive (IRSA) which is operated by the Jet Propulsion Laboratory, California Institute of  Technology, under contract with the National Aeronautics and Space  Administration.
\end{acknowledgements}

\bibliographystyle{aa}
\bibliography{ASAS-RCB-South}


\end{document}